\documentclass[twocolumn,tighten,times,twocolappendix]{aastex631}

\begin{document}
\newcommand{\oi}{\text{[\ion{O}{1}]}}
\newcommand{\oii}{\text{[\ion{O}{2}]}}
\newcommand{\neiii}{\text{[\ion{Ne}{3}]}}
\newcommand{\oiii}{\text{[\ion{O}{3}]}}
\newcommand{\woiii}{\text{$W_\lambda(\oiii)$}}
\newcommand{\nii}{\text{[\ion{N}{2}]}}
\newcommand{\hei}{\text{\ion{He}{1}}}
\newcommand{\heii}{\text{\ion{He}{2}}}
\newcommand{\ha}{\text{H$\alpha$}}
\newcommand{\wha}{\text{$W_\lambda(\ha)$}}
\newcommand{\hb}{\text{H$\beta$}}
\newcommand{\hg}{\text{H$\gamma$}}
\newcommand{\hd}{\text{H$\delta$}}
\newcommand{\he}{\text{H$\epsilon$}}
\newcommand{\hz}{\text{H$\zeta$}}
\newcommand{\hn}{\text{H$\eta$}}
\newcommand{\htheta}{\text{H$\theta$}}
\newcommand{\hiota}{\text{H$\iota$}}
\newcommand{\pa}{\text{Pa$\alpha$}}
\newcommand{\pb}{\text{Pa$\beta$}}
\newcommand{\pg}{\text{Pa$\gamma$}}
\newcommand{\pd}{\text{Pa$\delta$}}
\newcommand{\hi}{\text{\ion{H}{1}}}
\newcommand{\hii}{\text{\ion{H}{2}}}
\newcommand{\hk}{\text{H$\kappa$}}
\newcommand{\caii}{\text{\ion{Ca}{2}}}
\newcommand{\sii}{\text{[\ion{S}{2}]}}
\newcommand{\siii}{\text{[\ion{S}{3}]}}
\newcommand{\wlya}{\text{$W_\lambda$({\rm Ly$\alpha$})}}
\newcommand{\wlyaem}{\text{$W_\lambda^{\rm em}$({\rm Ly$\alpha$})}}
\newcommand{\llya}{\text{$L$(Ly$\alpha$)}}
\newcommand{\llyaobs}{\text{$L$(Ly$\alpha$)$_{\rm obs}$}}
\newcommand{\llyaint}{\text{$L$(Ly$\alpha$)$_{\rm int}$}}
\newcommand{\lyafrac}{\text{$f_{\rm esc}^{\rm spec}$(Ly$\alpha$)}}
\newcommand{\lha}{\text{$L$(H$\alpha$)}}
\newcommand{\lhb}{\text{$L$(H$\beta$)}}
\newcommand{\sfrha}{\text{SFR(\ha)}}
\newcommand{\sfrneb}{{\rm SFR}_{\rm neb}}
\newcommand{\sfrsed}{\text{SFR(SED)}}
\newcommand{\sfruv}{\text{SFR(UV)}}
\newcommand{\ssfrha}{\text{sSFR(\ha)}}
\newcommand{\ssfrsed}{\text{sSFR(SED)}}
\newcommand{\ebmv}{E(B-V)}
\newcommand{\ebmvneb}{E(B-V)_{\rm neb}}
\newcommand{\ebmvcont}{E(B-V)_{\rm cont}}
\newcommand{\ebmvlos}{E(B-V)_{\rm los}}
\newcommand{\nhi}{N(\text{\ion{H}{1}})}
\newcommand{\lognhi}{\log[\nhi/{\rm cm}^{-2}]}
\newcommand{\lognhitable}{\log\left[\frac{\nhi}{{\rm cm}^{-2}}\right]}
\newcommand{\lya}{\text{Ly$\alpha$}}
\newcommand{\lyb}{\text{Ly$\beta$}}
\newcommand{\lyg}{\text{Ly$\gamma$}}
\newcommand{\comment}[1]{}
\newcommand{\wciii}{\text{$W_\lambda$(\ion{C}{3}])}}
\newcommand{\ciii}{\text{\ion{C}{3}]}}
\newcommand{\interoiii}{\text{\ion{O}{3}]}}
\newcommand{\rsiione}{R(\text{\ion{Si}{2}}\lambda 1260)}
\newcommand{\rsiitwo}{R(\text{\ion{Si}{2}}\lambda 1527)}
\newcommand{\rsii}{R(\text{\ion{S}{2}})}
\newcommand{\cii}{\text{\ion{C}{2}}}
\newcommand{\civ}{\text{\ion{C}{4}}}
\newcommand{\siiv}{\text{\ion{Si}{4}}}
\newcommand{\rcii}{R(\text{\ion{C}{2}}\lambda 1334)}
\newcommand{\ralii}{R(\ion{Al}{2}\lambda 1670)}
\newcommand{\qh}{Q(\text{H$^0$})}
\newcommand{\rs}{{\cal R}_{\rm s}}
\newcommand{\fcov}{f_{\rm cov}}
\newcommand{\fcovhi}{f_{\rm cov}(\hi)}
\newcommand{\fcovmetal}{f_{\rm cov}({\rm metal})}
\newcommand{\fesclya}{f_{\rm esc}^{\rm spec}(\lya)}
\newcommand{\logxi}{\log[\xi_{\rm ion}/{\rm s^{-1}/erg\,s^{-1}\,Hz^{-1}}]}
\newcommand{\logq}{\log[Q/{\rm s^{-1}}]}
\newcommand{\lir}{L_{\rm IR}}
\newcommand{\lbol}{L_{\rm bol}}
\newcommand{\luv}{L({\rm UV})}
\newcommand{\rv}{R_V}

\title{The AURORA Survey: Multiple Balmer and Paschen Emission Lines for Individual Star-forming Galaxies at $z=1.5-4.4$. II. Implications for Nebular Dust Corrections, Nebular SFRs, and Differential Reddening}

\author[0000-0001-9687-4973]{Naveen A. Reddy}
\affiliation{Department of Physics and Astronomy, University of California, 
Riverside, 900 University Avenue, Riverside, CA 92521, USA; naveenr@ucr.edu}

\author[0000-0003-3509-4855]{Alice E. Shapley}
\affiliation{Department of Physics \& Astronomy, University of California,
Los Angeles, 430 Portola Plaza, Los Angeles, CA 90095, USA}

\author[0000-0003-4792-9119]{Ryan L. Sanders}
\affiliation{Department of Physics, University of California, Davis, One Shields Ave, Davis, CA 95616, USA}

\author[0000-0001-8426-1141]{Michael W. Topping}
\affiliation{Steward Observatory, University of Arizona, 933 North 
Cherry Avenue, Tucson, AZ 85721, USA}

\author[0000-0002-5139-4359]{Max Pettini}\affiliation{Institute of Astronomy, Madingley Road, Cambridge CB3 OHA, UK}

\author[0000-0003-4264-3381]{Natascha M. F\"{o}rster Schreiber}\affiliation{Max-Planck-Institut f{\"u}r extraterrestrische Physik (MPE), Giessenbachstr. 1, D-85748 Garching, Germany}

\author[0000-0002-4834-7260]{Charles C. Steidel}\affiliation{Cahill Center for Astronomy and Astrophysics, California Institute of Technology, MS 249-17, Pasadena, CA 91125, USA}

\author[0000-0003-1249-6392]{Leonardo Clarke}\affiliation{Department of Physics \& Astronomy, University of California,
Los Angeles, 430 Portola Plaza, Los Angeles, CA 90095, USA}

\author[0000-0001-7782-7071]{Richard S. Ellis}\affiliation{Department of Physics \& Astronomy, University College London, Gower St., London WC1E 6BT, UK}

\author[0000-0003-4464-4505]{Anthony J. Pahl}
\altaffiliation{Carnegie Fellow}
\affiliation{The Observatories of the Carnegie Institution for Science, 813 Santa Barbara Street, Pasadena, CA 91101, USA}

\author[0000-0002-8096-2837]{Garth D. Illingworth}\affiliation{Department of Astronomy and Astrophysics, UCO/Lick Observatory, University of California, Santa Cruz, CA 95064, USA}

\author[0000-0002-7613-9872]{Mariska Kriek}\affiliation{Leiden Observatory, Leiden University, NL-2300 RA Leiden, Netherlands}

\author[0000-0002-7064-4309]{Desika Narayanan}\affiliation{Department of Astronomy, University of Florida, 211 Bryant Space Sciences Center, Gainesville, FL 32611 USA}
\affiliation{Cosmic Dawn Center at the Niels Bohr Institute, University of Copenhagen and DTU-Space, Technical University of Denmark}


\begin{abstract}

We discuss the implications of the nebular dust attenuation curves
derived for 24 galaxies at $z=1.5-4.4$ with multiple detections of
$\hi$ Balmer and Paschen recombination emission lines from the
JWST/AURORA survey. The total attenuation of $\ha$ is $\approx
0.20$\,dex larger on average than the values obtained with the
commonly adopted combination of the Balmer decrement and the Galactic
extinction curve.  Nebular-line SFRs ($\sfrneb$) and SED-based SFRs
($\sfrsed$) are consistent on average when using the MOSDEF
attenuation or SMC extinction curves for the latter.  The relation
between nebular and stellar reddening is consistent with a scenario
where the Paschen lines are sensitive to heavily reddened OB
associations, while relatively unreddened OB associations contribute
significantly to the Balmer line and UV continuum emission.  There is
a stark contrast between the low $\rv$ (or steepness) of stellar dust
attenuation curves and the high $\rv$ (or flatness) of nebular dust
attenuation curves.  We suggest that the latter could be reflective of
a more porous medium established by strong feedback from massive
stars.  For the youngest galaxy in the sample, the stellar reddening
curve is identical to the nebular attenuation curve, in accordance
with our expectation that OB associations dominate the stellar
continuum emission at all wavelengths for this very young galaxy.
Larger samples will be needed to determine whether this conclusion
holds for young galaxies in general, and provide further insights into
the dust and metals mixing timescale on the scale of $\hii$ regions.

\end{abstract}


\section{Introduction}
\label{sec:intro}

The near-IR wavelength coverage and sensitivity provided by the James
Webb Space Telescope (JWST) have led to transformative advances in the
characterization of the stellar populations and interstellar medium
(ISM) of galaxies across a wide range of luminosity and redshift.
Determining key properties of the ISM, such as ionization parameter
and gas-phase abundance, and total nebular star-formation rates
(SFRs), often rely on measurements of dust-corrected line ratios and
luminosities.  Thus, accurate constraints on the nebular dust
attenuation curves, dust reddening, and total attenuation are crucial
for determining these properties of the ISM and total SFRs.  In
addition to $\hi$ Balmer recombination emission lines, JWST enables
the routine measurement of rest-frame near-IR $\hi$ Paschen lines for
star-forming galaxies, which were previously inaccessible or too faint
to detect.  When combined with the $\hi$ Balmer recombination emission
lines, the Paschen lines provide improved constraints on the
normalization and shape of the nebular dust attenuation curve
\citep{prescott22, sanders25} and nebular reddening (e.g.,
\citealt{reddy23a, sanders25}).  In \citet{reddy26a}, hereafter R26a,
we used deep JWST/NIRSpec spectra from the ``Assembly of Ultradeep
Observations Revealing Astrophysics'' (AURORA) survey
(PID:1914)---covering between 5 to 16 (median of 11) $\hi$ Balmer and
Paschen emission lines with $>5\sigma$ detections---to provide the
most robust constraints on the nebular attenuation curves of 24
star-forming galaxies at redshifts $z=1.4-4.4$.

We found that the reddening deduced from the Paschen lines is
systematically larger than that inferred from the Balmer lines under
the commonly-adopted assumptions of the Galactic extinction curve
\citep{cardelli89} and a $100\%$ covering fraction of dust.  This
result implies that some fraction of the star formation is heavily
reddened and does not contribute significantly to the bluer Balmer
lines.  Thus, line ratios formed from any combination of the Balmer
lines (e.g., the Balmer decrement, $\ha/\hb$) will not be particularly
sensitive to this dust-obscured star formation.  As a consequence, the
nebular attenuation curves of these galaxies---which account for
variations in dust optical depth---deviate from the commonly assumed
Galactic extinction curve, which represents a simple foreground screen
with unity dust covering fraction.  

Using the methodology first presented in \citet{reddy20}, we directly
calculated the effective nebular attenuation curves, $k_{\rm neb}^{\rm
  eff}(\lambda)$, using all available $\hi$ Balmer and Paschen lines
for each of the 24 galaxies.  The derived curves have a range of
normalizations, with the ratio of the total to selective extinction,
$\rv \equiv A_V/\ebmvneb$, varying in the range $3.2-16.4$ with most
having $\rv \ga 5$, significantly larger than the range of $\rv$ found
in various environments in the Milky Way ($\rv \simeq 2-6$)
\citep{cardelli89, fitzpatrick99, clayton00, gordon03, fitzpatrick09,
  zhang23}.  The offsets in reddening deduced from the Balmer and
Paschen emission lines, and the high $\rv$ values, reflect the
importance of the geometry of dust and stars in shaping the nebular
attenuation curves.

Along these lines, we proposed a model in which varying dust optical
depth within a galaxy is parameterized by the covering fraction of
dust, in addition to the nebular reddening.  In this model, the
observed line flux is a combination of light that escapes unimpeded by
dust and light that is subject to dust reddening.  Thus, the observed
flux as a function of wavelength can be written as follows:
\begin{eqnarray}
f(\lambda) & = & (1-\fcov)f_{0}(\lambda) + \nonumber \\
& & \fcov f_{0}(\lambda)\times 10^{-0.4\ebmvneb^{\rm cov}k_{\rm neb}(\lambda)}.
\label{eq:subunityeq1}
\end{eqnarray}
Here, $\fcov$ is the covering fraction of dust, $f_0(\lambda)$ is the
intrinsic flux at wavelength $\lambda$, $\ebmvneb^{\rm cov}$ is the
reddening towards the dust-covered regions of the galaxy, and $k_{\rm
  neb}(\lambda)$ is the reddening curve, typically assumed to be the
Galactic extinction curve \citep{cardelli89}.  Fitting this model to
all available $\hi$ Balmer and Paschen lines implies $\fcov \sim
0.6-1.0$ for a Galactic extinction curve.  Thus, the effective nebular
attenuation curves can be reproduced by assuming the Galactic
extinction curve with a subunity covering fraction of dust.  We
further found that the $\rv$ of the effective nebular attenuation
curves are primarily driven by $\fcov$ and the mix of reddened and
unreddened OB associations: galaxies with high $\rv$ have a larger
fraction of unreddened light contributing to the line fluxes.  Further
details on the supporting analysis and results are presented in R26a.

The evidence for sub-unity covering fractions of dust towards OB
associations, as discussed in R26a (see also \citealt{prescott22,
  wozniak26}), has direct implications for the relative reddening of
the nebular lines and stellar continuum (differential reddening).  In
the traditional interpretation, the ionizing flux giving rise to the
$\hi$ recombination lines is dominated by very massive O stars ($\ga
8$\,$M_\odot$) with lifetimes shorter than the typical molecular cloud
crossing timescale of $\la 20$\,Myr \citep{blitz80, calzetti94,
  mckee07}.  Once these stars explode as supernovae and the molecular
cloud is disrupted, they will no longer contribute substantially to
the ionizing flux.  OB associations that survive longer than the cloud
dissipation timescale will no longer contain the most massive O stars
and will not dominate the ionizing flux, but will still contribute
appreciably to the far-UV non-ionizing stellar continuum.  Thus, the
far-UV continuum (and stellar continuum in general) is expected to
suffer less attenuation compared to the nebular lines at a given
wavelength, an expectation that appears to be borne out by
observations of both nearby and distant star-forming galaxies
\citep{fanelli88, calzetti97, calzetti00, forster09, wuyts11,
  kashino13, kreckel13, price14, reddy15, debarros16, buat18,
  shivaei20b, reddy20, fetherolf21, lorenz23, lorenz24}.

However, reddening differences that are seen even between the Balmer
and Paschen lines when assuming a $100\%$ covering fraction of
dust---these lines originate from the same $\hii$ regions around the
same massive stars---imply variations in dust optical depth towards OB
associations that may be unrelated to their lifetimes.  In the case of
the youngest galaxies ($\la 10$\,Myr) where the stellar continuum at
all wavelengths is dominated by OB stars, we expect that the stellar
and nebular reddening curves should be similar (e.g.,
\citealt{reddy20}).  The AURORA sample---with measurements of Balmer
and Paschen lines, and a wealth of multi-wavelength photometry to
constrain the reddening of the stellar continuum---can be used to shed
light on these differences in nebular reddening and, by extension, the
differences between nebular and stellar reddening.  Here, we extend
upon the analysis presented in R26a by examining the impact of the
nebular attenuation curves on dust-corrected line ratios and
luminosities, nebular SFRs, and differential reddening.  We refer the
reader to \citet{pahl26} for further discussion of how these nebular
attenuation curves affect other quantities of interest, specifically
the ionizing photon production efficiency, $\xi_{\rm ion}$.

This paper is organized as follows.  The AURORA survey, data reduction
and measurements, and sample selection are briefly discussed in
Section~\ref{sec:data}.  The impact of the individual nebular dust
attenuation curves on reddening and total attenuation, and the
implications for dust-corrected line luminosities and ratios, are
presented in Section~\ref{sec:dustcorrect}.  Revised nebular SFRs,
their comparison to SED-inferred SFRs, and the implications for the
predicted relationship between dust attenuation and UV slope are
described in Section~\ref{sec:sfrs}.  This section also discusses the
impact of the revised nebular SFRs on the relationship between SFR and
$M^\ast$, and the consequences for calibrating JWST/MIRI-based SFRs.
The differential reddening of the stellar continuum and nebular
emission, and the convergence of the nebular and stellar attenuation
curves for young galaxies, are presented in
Section~\ref{sec:differential}.  A \citet{chabrier03} initial mass
function (IMF) is considered throughout the paper.  Wavelengths are
reported in the vacuum frame.  We adopt a cosmology with
$H_{0}=70$\,km\,s$^{-1}$\,Mpc$^{-1}$, $\Omega_{\Lambda}=0.7$, and
$\Omega_{\rm m}=0.3$.  A solar mass fraction of metals of $Z_\odot =
0.0142$ \citep{asplund09} is adopted.

\section{Data, Measurements, and Sample Selection}
\label{sec:data}

The JWST/NIRSpec Cycle 1 AURORA program (PID:1914, co-PIs: Shapley and
Sanders) obtained 12.3, 8.0, and 4.2-hr depth spectroscopy in the
G140M/F100LP, G235M/F170LP, and G395M/F290LP grating/filter
combinations, respectively, for 97 galaxies at redshifts $z=1.4-10.4$
in the GOODS-N and COSMOS fields.  This instrumental setup and the
depths yield a uniform $3\sigma$ line flux detection limit of $5\times
10^{-19}$\,erg\,s$^{-1}$\,cm$^{-2}$ from 1 to 5\,$\mu$m.

The standard STScI NIRSpec data reduction pipeline and custom software
developed for the AURORA survey were used to process the raw data to
produce fully calibrated two-dimensional (2D) spectrograms.
One-dimensional spectra were generated using optimal extraction.  Slit
loss corrections were performed by smoothing the F115W images of the
galaxies to reproduce their light profiles at longer wavelengths, and
then calculating the fraction of light passing through the
microshutter aperture.  Typical values for the transmitted fraction at
$2.7$\,$\mu$m are $20-60\%$.  The spectra were divided by these
wavelength-dependent transmission fractions, and an overall shift in
flux was implemented to ensure consistency with the broadband
photometry.  Finally, line fluxes were measured by fitting Gaussian
function(s) to the emission lines and assuming a continuum defined by
the best-fit SEDs to the photometry.  The latter allows us to account
for underlying stellar absorption when measuring the $\hi$
recombination emission-line fluxes.  These initial flux measurements
were used to correct the broadband photometry for the emission lines.
The SEDs were then refit to the corrected photometry, and the line
fluxes were remeasured using the resulting best-fit SEDs.  

The SED fitting uses all publicly available photometry from the
optical through the near-IR with JWST/NIRCam and HST/ACS and WFC3, as
obtained from the Dawn JWST Archive (DJA;
\citealt{heintz25}).\footnote{https://dawn-cph.github.io/dja/index.html}
The Binary Population and Spectral Synthesis (BPASS) version 2.2.1
models \citep{eldridge17, stanway18} were used to fit the photometry.
We considered constant star-formation models that include the effects
of binary stellar evolution, ages ranging from 1 Myr to the age of the
Universe at the redshift of each galaxy; a {\em stellar} metallicity
of $Z_\ast = 0.001$ (corresponding to $Z_\ast \simeq 1/10 Z_\odot$;
e.g., see \citealt{steidel16, cullen19, topping20a, topping20b,
  reddy22}); and a 100\,$M_\odot$ cutoff of the IMF.  The fiducial
modeling assumes the SMC extinction curve for the reddening of the
stellar continuum, but other stellar reddening curves are considered
in Section~\ref{sec:sfrs}.  Survey details, data reduction methods,
slitloss corrections and flux calibrations, line measurements, and the
SED fitting are described in more detail in \citet{sanders25},
\citet{shapley25}, \citet{topping25}, and in R26a.

Of the 94 galaxies with spectroscopic redshifts in the parent AURORA
sample, 24 were included in the analysis of the nebular attenuation
curves.  The primary selection criteria for the 24 objects includes
the absence of a significant AGN component; $>5\sigma$ detections of
at least 5 lines, two of which must be Paschen lines; and
Balmer-line-based reddening that is $>3\sigma$ larger than zero.
Further details on the sample selection are provided in R26a.  The
resulting sample includes galaxies over the redshift range $z=1.5-4.4$
with SFRs$\simeq 1 - 500$\,$M_\odot$\,yr$^{-1}$ and stellar masses
$M_\ast \simeq 2\times 10^8 - 3\times 10^{10}$\,$M_\odot$.

\begin{figure}
  \epsscale{1.15}
  \includegraphics[width=1.0\linewidth]{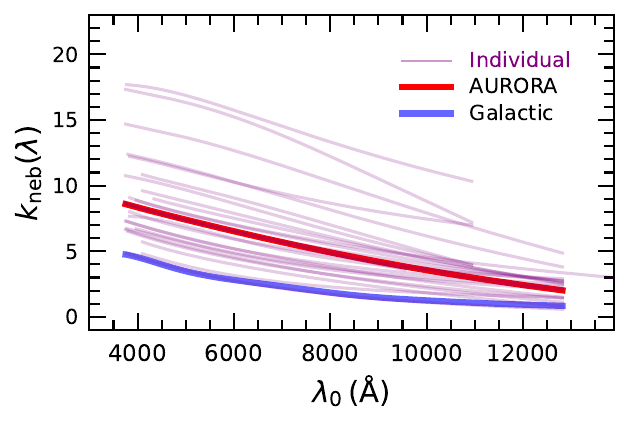}
    \caption{The average nebular attenuation curve, $k_{\rm
        neb}(\lambda)$, versus $\lambda_0$ for the main sample of 24
      galaxies (thick red curve).  The individual nebular dust
      attenuation curves are indicated by the light purple lines.  The
      Galactic extinction curve \citep{cardelli89} is shown by the
      solid blue line.}
   \label{fig:ktotalp2}
\end{figure}

\begin{figure}
  \epsscale{1.10}
  \includegraphics[width=1.0\linewidth]{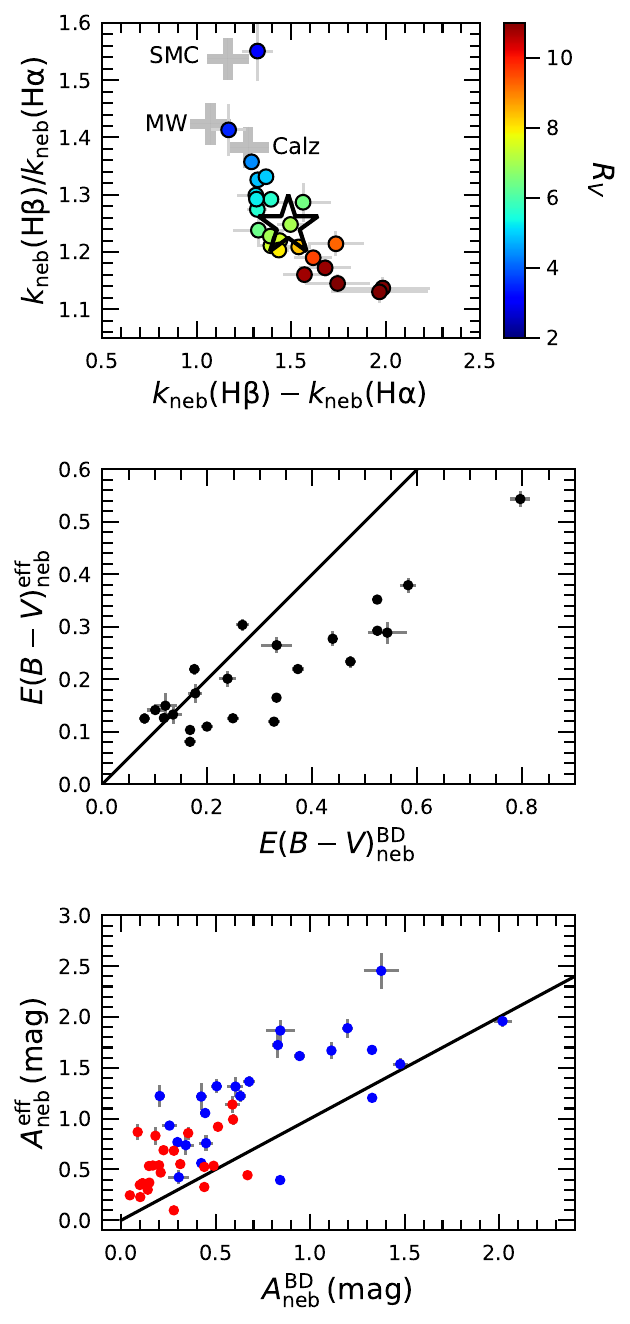}
    \caption{{\em Top:} Distribution of the attenuation curve slopes
      between $\hb$ and $\ha$, or $k_{\rm neb}^{\rm eff}(\hb)/k_{\rm
        neb}^{\rm eff}(\ha)$, and the attenuation curve differences
      between $\hb$ and $\ha$, or $k_{\rm neb}^{\rm eff}(\hb) - k_{\rm
        neb}^{\rm eff}(\ha)$.  The points are color-coded by $\rv$.
      Values for the average nebular dust attenuation curve are
      indicated by the open star.  The large grey crosses indicate
the difference and slope values for the Galactic extinction curve (MW),
the SMC extinction curve, and the Calzetti attenuation curve.
{\em Middle:} Comparison of the reddening derived using the
      effective attenuation curves from AURORA and those computed
      using the Galactic extinction curve.  The line of equality is
      indicated in black.  {\em Bottom:} Comparison of the attenuation
      derived using the effective attenuation curve and those computed
      using the Galactic extinction curve for $\ha$ (in blue) and the
      longest-wavelength Paschen line available for each galaxy (in
      red).  The line of equality is indicated in black.}
   \label{fig:bdcompare}
\end{figure}

\section{Implications for Dust-corrected Line Luminosities and Ratios}
\label{sec:dustcorrect}

In R26a, we calculated the individual nebular dust attenuation
curves for each of the 24 galaxies in the main sample, along with the
average of these curves, which are reproduced in
Figure~\ref{fig:ktotalp2} for convenience.  The average nebular attenuation curve
for the 24 galaxies in the sample is given by the following equation:
\begin{eqnarray}
\langle k_{\rm neb}(\lambda)\rangle & = & -7.241 +
\frac{17.002}{\lambda/\mu{\rm m}} - \frac{8.086}{(\lambda/\mu{\rm
    m})^2} \nonumber \\ 
& & + \frac{2.177}{(\lambda/\mu{\rm m})^3} -
\frac{0.319}{(\lambda/\mu{\rm m})^4} + \frac{0.021}{(\lambda/\mu{\rm
    m})^5},
\label{eq:avecurve}
\end{eqnarray}
valid over the wavelength range $0.35\la \lambda \la 1.28$\,$\mu$m.
Here, we assess the impact of both the individual and average nebular
attenuation curves on dust-corrected line luminosities and line
ratios, which are commonly used to infer SFRs and other important
physical properties of the ISM (e.g., ionization parameters, gas-phase
abundances).  This assessment is done in comparison to the standard
assumptions used in previous studies, which lacked direct constraints
on the shape and normalization of the nebular attenuation curve.

Before exploring these issues, it is important to evaluate the impact
of the individual nebular attenuation curves on the reddening and
total attenuation, relative to the values obtained with the commonly
adopted method of combining the $\ha/\hb$ ratio (i.e., the Balmer
decrement) with the Galactic extinction curve
(Section~\ref{sec:bdcompare}).
Sections~\ref{sec:effectonlineluminosities} and
\ref{sec:effectonlineratios} present a comparison of the
dust-corrected line luminosities and line ratios that assume this
commonly-adopted approach, versus those that assume the individual
nebular attenuation curves.  Section~\ref{sec:biaseswithave} discusses
the scatter induced in the nebular reddening and dust-corrected line
luminosities and line ratios when adopting an average nebular
attenuation curve in lieu of individual nebular attenuation curves.

\subsection{Comparison with Balmer-decrement-inferred Reddening and Attenuation}
\label{sec:bdcompare}

For the subsequent discussion, we adopt the traditional definition of
the ``slope'' of the nebular attenuation curve as the ratio of the
curve at two wavelengths: $k_{\rm neb}^{\rm eff}(\lambda_1)/k_{\rm
  neb}^{\rm eff}(\lambda_2)$.  Thus, the slope of the attenuation
curve between $\ha$ and $\hb$ is $k_{\rm neb}^{\rm eff}(\hb)/k_{\rm
  neb}^{\rm eff}(\ha)$.  The most common method of correcting for
nebular attenuation is to calculate the Balmer decrement, typically
the $\ha/\hb$ ratio, and assume the Galactic extinction curve to
calculate the reddening and attenuation.  The top panel of
Figure~\ref{fig:bdcompare} shows the distribution of attenuation curve
slopes and differences between $\hb$ and $\ha$: i.e., $k_{\rm
  neb}^{\rm eff}(\hb)/k_{\rm neb}^{\rm eff}(\ha)$ and $k_{\rm
  neb}^{\rm eff}(\hb) - k_{\rm neb}^{\rm eff}(\ha)$.  For almost all
the galaxies in the sample, the slope of the attenuation curve between
$\hb$ and $\ha$ is shallower (or greyer) than that of the Galactic
extinction curve.  The color coding of points in the top panel of
Figure~\ref{fig:bdcompare} demonstrates that galaxies with shallower
slopes between $\ha$ and $\hb$ have higher $\rv$, consistent with the
usual interpretation of higher $\rv$ corresponding to flatter or
greyer attenuation curves (e.g., \citealt{salim20}).

The top panel of Figure~\ref{fig:bdcompare} shows that the {\em
  differences} in the attenuation curves between $\hb$ and $\ha$ are
systematically {\em larger} than that of the Galactic extinction
curve.  $\ebmvneb$ is related to the difference in the attenuation curves
between $\hb$ and $\ha$ by the following equation:
\begin{equation}
\ebmvneb = \frac{2.5}{k_{\rm neb}^{\rm eff}(\hb) - k_{\rm neb}^{\rm eff}(\ha)}\log_{10}\left(\frac{\ha/\hb}{2.79}\right),
\end{equation}
where 2.79 is the intrinsic $\ha/\hb$ ratio assumed in R26a,
corresponding to an electron density of $n_e = 100$\,cm$^{-3}$ and
electron temperature of $T_e = 15,000$\,K.  The $\ebmvneb$ calculated
based on the effective attenuation curves will be smaller than those
derived with the Galactic extinction curve.  In other words, the
larger difference in the attenuation curve at $\hb$ and $\ha$ implies
that less reddening is required to reproduce the observed line ratios.
The same conclusions hold true with respect to the SMC and
\citet{calzetti00} curves.  The middle panel of
Figure~\ref{fig:bdcompare} shows that for a majority of the galaxies
in the sample, $\ebmvneb$ computed with the effective attenuation
curve are typically a factor of $1.5\times$ lower than the $\ebmvneb$
computed based on the Balmer decrement and the Galactic extinction
curve.  Thus, while the attenuation curve is expected to show the
greatest variation at UV wavelengths, our analysis indicates that
there is still enough variation at optical wavelengths to produce a
factor of $\simeq 1.5$ change in reddening at a fixed Balmer
decrement.

Despite the lower $\ebmvneb^{\rm eff}$, the high $\rv$ of the galaxies
results in a total attenuation (e.g., of $\ha$, or the
longest-wavelength available Paschen line for each galaxy) that is
generally larger than that computed from the Galactic extinction curve
(bottom panel of Figure~\ref{fig:bdcompare})---typically $0.52$\,mag
larger at $\ha$ (see also \citealt{pahl26}) and $\simeq 0.27$\,mag
larger at the longest-wavelength Paschen line available for each
galaxy.  Here, $A_{\rm neb}^{BD}$ was computed by inferring the
$\ebmvneb$ from the Balmer decrement and the Galactic extinction
curve, and then using the Galactic extinction curve along with the
$\ebmvneb$ to compute the attenuation at the wavelength of $\ha$ and
the longest-wavelength Paschen line.  An important conclusion from our
analysis is that even the relatively long-wavelength Pa5 (Pa$\beta$)
line ($\lambda_0 = 1.28$\,$\mu$m) suffers a non-negligible attenuation
of $0.11$\,mag on average, even among the modestly dust-reddened
galaxies with $\log[M^\ast/M_\odot] = 8.5-10.5$ in our sample.

\subsection{Effect on Line Luminosities}
\label{sec:effectonlineluminosities}

The impact of the individual nebular attenuation curves on line
luminosities is demonstrated by focusing on a few of the strong
nebular emission lines across a broad wavelength range, in particular
$\oii\lambda\lambda 3727,3730$, $\hb$, $\oiii\lambda\lambda
4960,5008$, $\ha$, $\sii\lambda\lambda 6718, 6733$, and
$\siii\lambda\lambda 9071, 9533$.  
Line luminosities corrected for dust with the individual effective
attenuation curves ($L_{\rm eff}$) were compared with those corrected
using the Balmer decrement and the Galactic extinction curve ($L_{\rm
  BD}$).  Figure~\ref{fig:lumcompare} in Appendix~\ref{sec:app1}
summarizes this comparison.

In general, $L_{\rm eff}$ is systematically larger than $L_{\rm BD}$
due to the larger total attenuation implied by the individual nebular
attenuation curves relative to that implied by the Galactic extinction
curve.  The values of $\log L_{\rm eff}$ are anywhere from 0.13 to
0.22\,dex larger than $\log L_{\rm BD}$ and, for $\ha$, the mean
offset is $0.20$\,dex.  For the longest-wavelength Paschen line
available for each galaxy, the mean offset is expectedly lower at
$0.11$\,dex.  As discussed further in Appendix~\ref{sec:app1}, these
mean offsets may be larger or smaller than the values quoted here,
depending on the reddening distribution of galaxies relative to that
of the AURORA sample.

\subsection{Effect on Line Ratios}
\label{sec:effectonlineratios}

We evaluated the effect of the individual nebular attenuation curves
on dust-corrected line ratios by focusing on O32, R23, and S32,
defined as follows:
\begin{eqnarray}
{\rm O32} & \equiv & \log(\oiii\lambda\lambda 4960,5008/\oii\lambda\lambda 3727,3730),
\label{eq:o32}
\end{eqnarray}
\begin{eqnarray}
{\rm R23} & \equiv & \log((\oiii\lambda\lambda 4960,5008\, + \oii\lambda\lambda 3727,3730)/\hb),
\label{eq:r23}
\end{eqnarray}
and
\begin{eqnarray}
{\rm S32} & \equiv & \log(\siii\lambda\lambda 9071,9533/\sii\lambda\lambda 6718,6733).
\label{eq:s32}
\end{eqnarray}
The O32 and S32 ratios are commonly used to probe the ionization state
of the ISM, while R23 is commonly used as an oxygen abundance
indicator.  Additionally, we considered two ratios involving the
auroral lines: $\oiii\lambda 4364/\oiii\lambda 5008$ and
$\oii\lambda\lambda 7320, 7330/\oii\lambda\lambda 3727,3730$, which
are used to deduce electron temperatures and obtain so-called
``direct'' oxygen abundances.  All of these ratios involve lines that
are widely separated in wavelength and will therefore suffer different
amounts of total attenuation for a given reddening.
Appendix~\ref{sec:app1} summarizes the comparison between line ratios
derived using the aforementioned two methods of correcting for dust.

The average
offset in the O32 line ratios computed with the two methodologies is
on average 0.052\,dex, implying $\approx 13\%$ larger O32 (and
inferred ionization parameter, $U$; see \citealt{reddy23b}) on average
when adopting the individual nebular attenuation curves.  The average
offset is smaller for galaxies with the largest O32 and R23 ratios.
These galaxies have higher ionization parameters, lower gas-phase
abundances, and lower dust attenuation, relative to galaxies with
smaller O32 and R23 ratios.
Not surprisingly, we find that the average shifts in the line ratios
are smaller than the average shifts in the absolute line luminosities
(see previous section).

For the line ratios involving the auroral lines, the mean offsets
assuming the individual nebular attenuation curves versus the Balmer
decrement and Galactic extinction curve are $-0.032$\,dex and
$0.066$\,dex, respectively, for the $\oiii\lambda 4364/\oiii\lambda
5008$ and $\oii\lambda\lambda 7320, 7330/\oii\lambda\lambda 3727,3730$
ratios.  The latter combines a blue rest-optical doublet and near-IR
doublet with the widest wavelength separation of $\Delta\lambda\approx
3600$\,\AA.  The offset in the ratio determined using the effective
attenuation curve and the Galactic extinction curve is as high as
$0.19$\,dex for individual objects.  Such a bias results in anywhere
from a $\simeq 20-50\%$ increase in the electron temperature ($T_e$)
inferred for the [OII]-emitting zone, while the impact on the derived
oxygen abundance is $\la 0.15$\,dex.  Thus, systematics in the assumed
nebular dust attenuation curve are a subdominant source of uncertainty
in oxygen abundance measurements, at least for the typical galaxy in
the AURORA sample.

\begin{figure*}
  \epsscale{1.00}
  \includegraphics[width=1.0\linewidth]{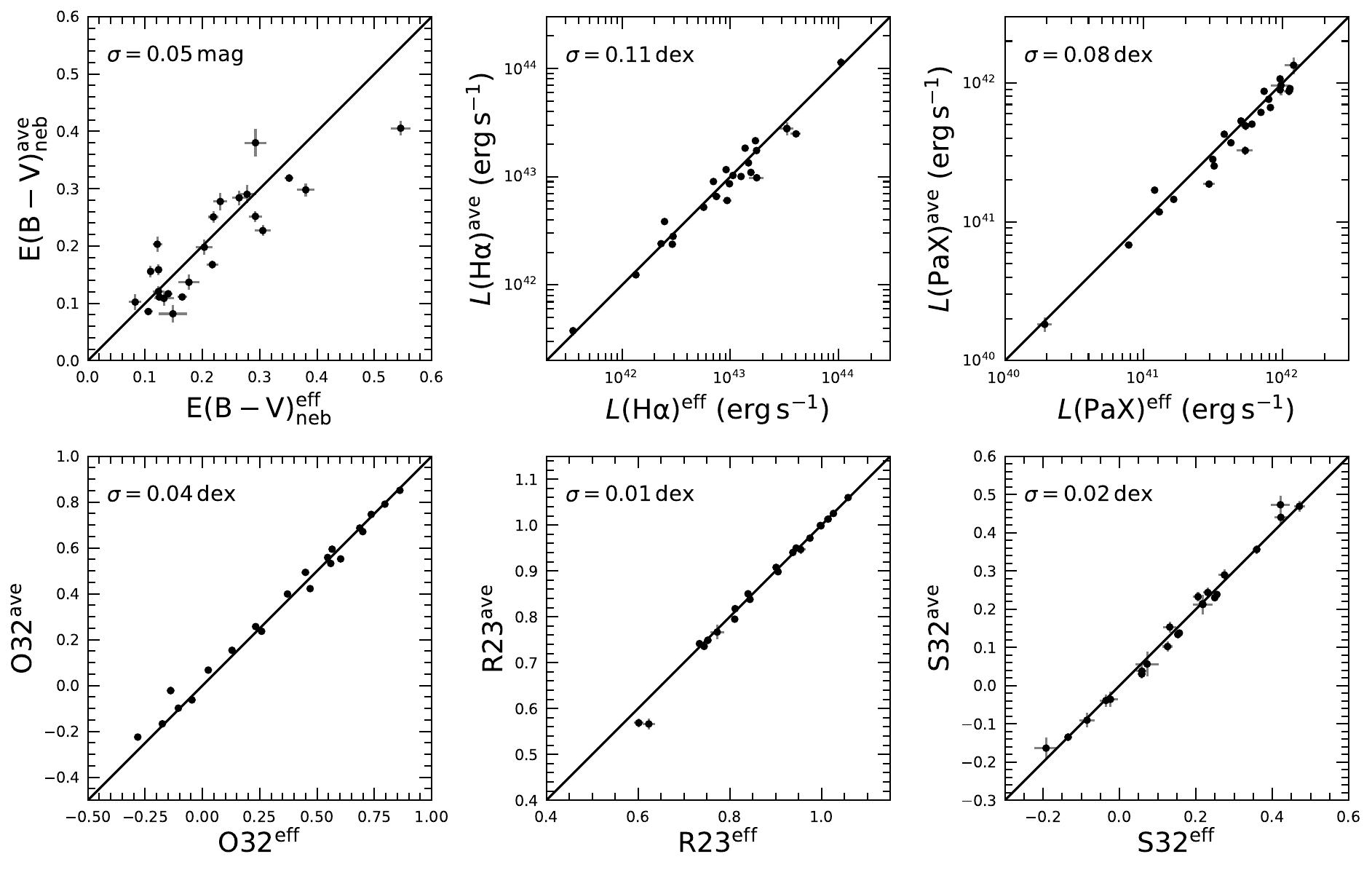}
    \caption{Comparison of $\ebmvneb$ and dust-corrected line
      luminosities ($\ha$ and the longest-wavelength Paschen line
      available for each galaxy) and line ratios (O32, R23, and S32)
      assuming the average and individual nebular attenuation curves.
      The standard deviation between the quantities assuming the
      average and individual nebular attenuation curves and indicated
      in each panel.}
   \label{fig:avevsind}
\end{figure*}

\subsection{Biases in Assuming the Average Nebular Attenuation Curve for Individual Galaxies}
\label{sec:biaseswithave}

Robust constraints on the nebular dust attenuation curve may not be
possible in cases where a significant fraction of the $\hi$
recombination lines are too weak to detect (e.g., for galaxies with
very low SFRs) or lack wavelength coverage (e.g., for galaxies at
$z\ga 5$ where all the Paschen lines fall out of the spectral coverage
of JWST/NIRSpec).  In such cases, one is often forced to rely on an
assumed average nebular dust attenuation curve to dust correct line
luminosities and ratios.  The assumption of an average curve is a
useful approximation---particularly when applied to statistical
samples of galaxies where variations in the dust/star geometry (or
dust properties) may average out.  Obviously, however, this assumption
is not ideal.  The factors that cause changes in the shape and/or
normalization of the curve---e.g., the geometry/covering fraction of
dust relative to stars---may vary considerably from galaxy to galaxy.
Thus, assuming an average dust curve will introduce uncertainty in the
dust corrections for individual galaxies, especially those that
deviate from the typical properties of the sample from which the
average curve is derived.

Figure~\ref{fig:avevsind} summarizes key results from comparing
dust-corrected quantities based on these two assumptions: individual
nebular attenuation curves versus the average nebular attenuation
curve (see also \citealt{pahl26}).  Each panel of the figure shows the
standard deviation of the difference in dust-corrected quantities,
depending on whether the individual or average nebular attenuation
curve is assumed.

For instance, the top-left panel demonstrates that assuming the
average nebular attenuation curve results in reddening values that
scatter around the ``true'' values obtained using the individually
determined nebular attenuation curves, with a standard deviation of
$\sigma = 0.05$\,mag. 
The
scatter in other line luminosities and line ratios is also indicated
in the figure. When using the average nebular attenuation curve, the
dust-corrected $\log[L(\ha)/{\rm erg\,s^{-1}}]$ can scatter by more
than $0.11$\,dex from the true values for $32\%$ of galaxies. For the
longest-wavelength Paschen line available for each galaxy, the scatter
is slightly smaller, with $\sigma = 0.08$\,dex.  In general, the line
ratios O32, R23, and S32 are less sensitive to the assumed dust
attenuation curve (i.e., either the individually determined nebular
attenuation curve or the average attenuation curve), exhibiting a
scatter of between $\sigma = 0.01-0.04$\,dex.  For the two
dust-corrected auroral line ratios that are not shown in
Figure~\ref{fig:avevsind}, $\oiii\lambda 4364/\oiii\lambda 5008$ and
$\oii\lambda\lambda 7320, 7330/\oii\lambda\lambda 3727,3730$, we find
$\sigma=0.01$ and $\sigma=0.05$\,dex, respectively.

The $\sigma$ values shown in Figure~\ref{fig:avevsind} provide an
important estimate of the additional uncertainty (or scatter) in
dust-corrected quantities introduced by assuming the average nebular
dust attenuation curve for individual galaxies.  Note that the
uncertainties may be larger for certain types of galaxies.  The
comparisons mentioned above include galaxies that were also all used
in computing the average attenuation curve.  Galaxies with dust
covering fractions and/or reddening (and hence attenuation curves)
that are atypical of those in the AURORA sample may experience even
larger uncertainties when the average curve derived from the AURORA
sample is assumed.  On the other hand, recall that the sample was
specifically selected to include only galaxies with non-negligible
reddening.  Galaxies with little to no reddening, which were excluded
from the sample (Section~\ref{sec:data}), would naturally show smaller
discrepancies between dust-corrected quantities derived from the
average curve and those derived from their individually determined
curves.  Since the AURORA sample is generally representative of
typical star-forming galaxies at $z\sim 1.5-4.5$ \citep{shapley25},
the scatter in dust-corrected quantities discussed above should
provide reasonable estimates of the additional uncertainty introduced
when assuming the average nebular dust attenuation curve for
individual galaxies.

\section{Nebular SFRs}
\label{sec:sfrs}

Each $\hi$ recombination line---once corrected for dust with a curve
computed for the galaxy in question---provides a semi-independent
estimate of the ionizing photon rate, $Q({\rm H})$, and the nebular
SFR.\footnote{The dust corrections for each line depend on the nebular
  dust attenuation curve, which is constrained by all available lines.
  As a result, the dust corrections applied to one line are not
  entirely independent of those applied to other lines.  Consequently,
  the dust-corrected $Q({\rm H})$ and SFR derived from one line is not
  fully independent of the dust-corrected $Q({\rm H})$ and SFR derived
  from a different line.}  Each galaxy in the sample has anywhere from
5 to 16 significantly detected $\hi$ lines, thus providing the same
number of (close to) independent estimates of $Q({\rm H})$ and the
nebular SFR.

Following \citet{leitherer95}, each dust-corrected $\hi$ recombination
line was converted to an SFR using the following equation:
\begin{eqnarray}
\sfrneb(M_\odot\,{\rm yr}^{-1}) & = & L(\lambda)({\rm erg\,s^{-1}})\left(\frac{\lha}{L(\lambda)}\right)_0 \frac{10^{12}}{1.36N({\rm H})},
\end{eqnarray}
where $L(\lambda)$ is the dust-corrected recombination line
luminosity, $(\lha/L(\lambda))_0$ is the intrinsic line ratio of $\ha$
to the specific recombination line (see Table~1 of R26a), and
$N({\rm H})$ is the ionizing photon rate in s$^{-1}$ per
$M_\odot$\,yr$^{-1}$ of star formation.  The ionizing photon rate was
determined from the stellar population synthesis model that best fits
the broadband photometry.  In general, $\log[N({\rm H})/{\rm s}^{-1}]
\approx 53.54$, but can be up to $\approx 0.3$\,dex lower for the
youngest galaxies in the sample with ages $\la 10$\,Myr.  The
resulting factor used to convert $\ha$ luminosity to SFR is
$2.12\times 10^{-42}$\,$M_\odot$\,yr$^{-1}$\,${\rm erg}^{-1}$\,s for
most of the galaxies in the sample \citep{reddy22}, but is as high as
$4.50\times 10^{-42}$\,$M_\odot$\,yr$^{-1}$${\rm erg}^{-1}$\,s for the
youngest galaxy in the sample, GOODSN-17940, which has an inferred age
of $\simeq 2$\,Myr assuming the MOSDEF stellar dust attenuation curve
from \citet{reddy15}.\footnote{These conversion factors assume a
  \citet{chabrier03} IMF with an upper-mass cutoff of
  $100$\,$M_\odot$.}  Unless stated otherwise, the MOSDEF stellar
attenuation curve was used to estimate the age of the galaxy, and thus
$N({\rm H})$ and $\sfrneb$.
The average $\sfrneb$ for each galaxy was calculated by averaging the
SFRs computed from each dust-corrected $\hi$ recombination line.  

For context, Appendix~\ref{sec:app2} presents correlations between
$\sfrneb$ and the free parameters in the covering-fraction model,
including the line-of-sight reddening, $\ebmvneb^{\rm cov}$; the
covering fraction of dust, $\fcov$; and the fraction of the intrinsic
$\ha$ luminosity that is obscured by dust, i.e., $f_{\rm
  obsc}(\ha)/f_0(\ha)$.  Section~\ref{sec:sfrcompare} presents a
comparison of $\sfrneb$ and SFRs estimated from fitting the broadband
photometry of galaxies, $\sfrsed$.  An alternative way of interpreting
the data is to examine the correlation between dustiness and UV
spectral slope, $\beta$, which is presented in
Section~\ref{sec:irxbeta}.  Section~\ref{sec:sfrmstar} discusses the
so-called ``main sequence'' of star formation in light of the revised
values of $\sfrneb$.  The implications of our analysis for calibrating
the relationship between mid-IR luminosity (e.g., as measured with
JWST/MIRI) and SFR are discussed in Section~\ref{sec:miri}.

\subsection{Population-Averaged Nebular-Line and SED-inferred SFRs}
\label{sec:sfrcompare}

A substantial body of both numerical simulation and observational
studies suggests that the star-formation histories (SFHs) of
individual galaxies may be highly stochastic, particularly for
low-mass galaxies or those at high redshift (e.g., \citealt{weisz12,
  hopkins14, dominguez15, guo16, sparre17, faucher18, emami19, atek22,
  clarke24, clarke25}).  One commonly used probe of this stochasticity
is the ratio of the H-ionizing to non-ionizing UV continuum
luminosity.  The former is sensitive to star formation on timescales
of a few Myr, while the latter is sensitive to star formation on tens
of Myr to $\sim 100$\,Myr \citep{kennicutt94}.  Consequently, SFRs
derived from the nebular lines (e.g., $\ha$) and the UV continuum may
disagree if the assumed luminosity-to-SFR conversion factor does not
account for the age of the stellar population, a feature that is
commonly exploited to isolate galaxies that may be experiencing a
burst of star formation (e.g., \citealt{glazebrook99, meurer09,
  weisz12, emami19, atek22, rezaee23}).\footnote{SED-inferred SFRs are
  less sensitive to age-related variations in the luminosity-to-SFR
  conversion because the age of the stellar population is treated as a
  free parameter when fitting the SEDs.}  However, when examining a
statistical sample of galaxies, the stochastic variations in the SFHs
of individual galaxies are expected to average out, such that the mean
nebular-line SFR should be consistent with that derived from the UV
continuum.  The new measurements of $\sfrneb$ can be compared to those
derived from the UV continuum or from SED fitting, using different
assumptions for the stellar reddening curve.  This comparison allows
us to determine the stellar reddening curve that provides the best
overall agreement with the nebular SFRs (e.g., \citealt{reddy10,
  reddy15, theios19, shivaei20a, reddy22}).\footnote{We have assumed
  that the escape fraction of ionizing photons ($f_{\rm esc}$) is
  negligible based on the most direct constraints available on the
  average escape fraction of galaxies at redshifts similar to those of
  our sample, which imply typical $f_{\rm esc}\la 10\%$ (e.g.,
  \citealt{steidel18, pahl21, reddy22}).  Otherwise, an upward
  correction of the dust-corrected recombination line luminosities and
  $\sfrneb$ would be needed.}

\begin{figure*}
  \epsscale{1.1}
  \includegraphics[width=1.0\linewidth]{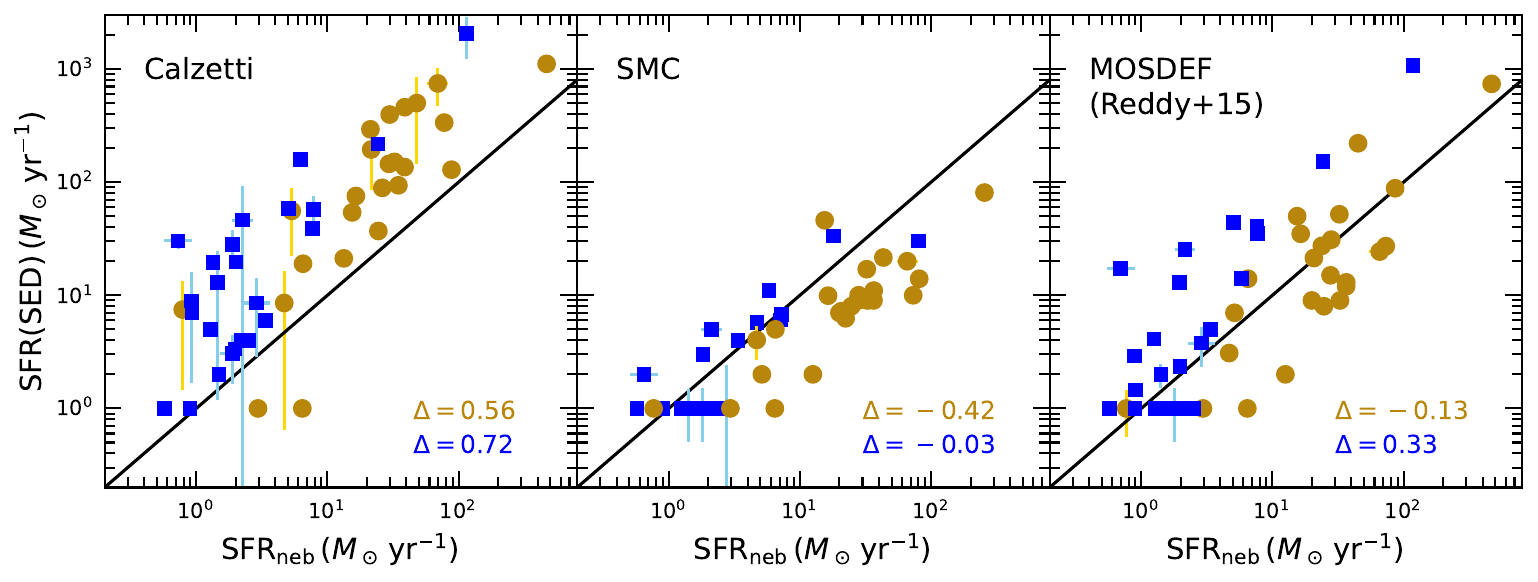}
    \caption{Comparison of $\sfrneb$ and SED-inferred SFR ($\sfrsed$)
      derived using the Calzetti (left), SMC (middle), and
      \citet{reddy15} (MOSDEF, right) dust attenuation curves.  The
      gold circles represent the 24 galaxies in the main sample, while
      the blue squares indicate galaxies excluded from the sample, but
      still with measured Balmer decrements.  For the excluded
      galaxies, $\sfrneb$ was computed assuming the Balmer decrement
      and the Galactic extinction curve (squares).
Each panel indicates the average offset in dex
      between $\log[\sfrsed/(M_\odot\,{\rm yr}^{-1})]$ and
      $\log[\sfrneb/(M_\odot\,{\rm yr}^{-1})]$ ($\Delta$) for the main
      sample of 24 galaxies and the excluded galaxies.  The best agreement between $\sfrneb$ and $\sfrsed$
      is achieved with the MOSDEF and SMC stellar reddening curves for
      the main and excluded samples, respectively, assuming a
      sub-solar stellar metallicity.}
   \label{fig:sfrcompare}
\end{figure*}

Figure~\ref{fig:sfrcompare} compares $\sfrneb$ with the SED-inferred
SFRs, $\sfrsed$, derived using three dust attenuation curves for the
stellar continuum: Calzetti, SMC, and the MOSDEF curve from
\citealt{reddy15} (see also \citealt{shivaei20a}).  The $\sfrneb$ are
computed self-consistently with the assumed SED model, in the sense
that the best-fit SED model for each assumed attenuation curve was
used to compute $N({\rm H})$ and subsequently the conversion factor
between nebular line luminosity and $\sfrneb$.  \footnote{The
  SED-inferred SFRs are comparable to those obtained by
  dust-correcting the UV luminosity using the UV spectral slope
  ($\beta$) along with the expected relationship between attenuation
  and UV slope for each of the stellar reddening curves mentioned
  above (e.g., \citealt{reddy18a}).}  Each panel indicates the average
offset in dex between $\log[\sfrsed/(M_\odot\,{\rm yr}^{-1})]$ and
$\log[\sfrneb/(M_\odot\,{\rm yr}^{-1})]$, denoted as $\Delta$.  For
the main sample of 24 galaxies used to derive the nebular dust
attenuation curves, the Calzetti and SMC curves systematically
overpredict ($\Delta=0.56$\,dex) and underpredict ($\Delta = -0.42$)
$\sfrneb$, respectively.  In contrast, the MOSDEF stellar reddening
curve from \citet{reddy15} provides the closest agreement between
$\sfrsed$ and $\sfrneb$, with $\Delta = -0.13$.

To determine how representative these results are, however, we must
account for sample selection.  As noted in Section~\ref{sec:data} (see
also Figure~4 of R26a), the main sample is biased towards galaxies
with slightly higher nebular reddening, as indicated by the Balmer
decrement, and higher observed $\ha$ luminosities compared to the
parent AURORA sample.  In other words, galaxies that were excluded
from the sample have lower dust attenuation and lower SFRs.  

These excluded galaxies consist of all star-forming objects in the
parent AURORA sample that had $>3\sigma$ detections of $\ha$ and
$\hb$, and at least one other $\hi$ recombination line.  There are 25
such galaxies in the excluded subsample.  Figure~\ref{fig:sfrcompare}
displays $\sfrsed$ and $\sfrneb$ for the excluded galaxies,
represented by squares, where $\sfrneb$ was estimated using the Balmer
decrement and the Galactic extinction curve.
For the excluded galaxies, the SMC reddening curve yields the best
agreement between $\sfrsed$ and the original Balmer-decrement-inferred
$\sfrneb$ ($\Delta = -0.03$).\footnote{If an upward correction of
  $0.20$\,dex is applied to $\sfrneb$ for the excluded
  galaxies---i.e., the same offset observed between $\sfrneb$ for the
  main sample of 24 galaxies when assuming the Balmer decrement and
  Galactic extinction curve versus the individual nebular attenuation
  curves---then the SMC and MOSDEF curves both provide the same level
  of agreement between $\sfrsed$ and $\sfrneb$ ($|\Delta| =
  0.13-0.23)$.  In contrast, the Calzetti curve results in the largest
  offsets between $\sfrsed$ and either the original or corrected
  $\sfrneb$ for the excluded galaxies ($\Delta > 0.52$).}

Recall that the SED fitting assumes the $Z_\ast = 0.001$ stellar
metallicity BPASS models (Section~\ref{sec:data}).  Generally,
the Calzetti attenuation curve leads to a better agreement between
$\sfrsed$ and $\sfrneb$ only when models with near-solar metallicity
($Z_\ast = 0.02$) are used.  This difference arises because the UV
slope ($\beta$) of the intrinsic stellar spectrum becomes redder with
increasing stellar metallicity, leading to lower inferred dust
attenuation and, consequently, lower $\sfrsed$.  As discussed in
\citet{shapley24}, SED modeling of the AURORA galaxies indicates that the 
combination of $\sim$solar-metallicity models and the Calzetti curve
generally yields a lower $\chi^2$ relative to the broadband photometry
than the combination of sub-solar-metallicity models and the SMC
curve.  That said,
multiple studies suggest that sub-solar stellar metallicities are more
appropriate for typical star-forming galaxies at $z\ga 2$ based on
stellar-population modeling of far-UV spectra (e.g.,
\citealt{steidel16, cullen19, cullen20, topping20a, topping20b,
  reddy22}).  Furthermore, a combined analysis of the O32 and S32
ratios of galaxies in the AURORA sample suggests that most of these
galaxies have hard ionizing spectra associated with
sub-solar-stellar-metallicity populations (Reddy et~al., in prep.).
In this context, our analysis suggests that the SMC and MOSDEF stellar
reddening curves provide the best match between $\sfrsed$ and
$\sfrneb$.  For comparison, an analysis of the Paschen-line-based SFRs
for $z\sim 1-3$ galaxies in the CEERS sample shows that these SFRs
align more closely with SMC-inferred $\sfrsed$ \citep{reddy23a}.  A
more refined comparison of SED- and nebular-based SFRs for galaxies
with constrained stellar metallicities will be presented elsewhere.

Whether the true shape of the average stellar reddening curve is
similar to that of the SMC or MOSDEF attenuation curves remains an
open question.  While direct constraints on the shape of the curve can
be obtained (e.g., \citealt{reddy15, shivaei20a}), the $\rv$ of these
curves ($\rv \simeq 2.5 - 3.1$) are lower than those typically
observed for nebular dust attenuation curves, a difference that is
likely tied to variations in optical depth to OB associations in the
galaxies (Section~\ref{sec:intro}; see also
Section~\ref{sec:rvcompare} for further discussion).  If these
variations in optical depth affect the contribution of reddened
emission to the optical Balmer lines, they will undoubtedly have a
more severe impact on the shorter-wavelength UV continuum emission
originating from the same OB associations.  The SED fitting does not
account for an optically thick component to the UV continuum, so the
agreement between $\sfrsed$ and $\sfrneb$---where only the latter
accounts for optically thick emission---may be coincidental.  Direct
measurements of the dust emission from these galaxies (e.g., via
JWST/MIRI or ALMA) can be used to constrain the dust-obscured UV (or
IR) luminosity.

\subsection{Inferred IRX-$\beta$ Relation}
\label{sec:irxbeta}

Along these lines, several analyses that employ direct measurements of
dust emission for $z\simeq 2 - 6$ galaxies suggest that the
relationship between dust obscuration---parameterized by the
infrared-to-unobscured UV luminosity, $\lir/\luv$ (or ``IRX'';
\citealt{meurer99})---and UV spectral slope ($\beta$) is largely
consistent with the one predicted by the SMC curve (e.g.,
\citealt{bouwens16a, reddy20, fudamoto20}), particularly for the
sub-solar stellar metallicities expected for these galaxies, though
with considerable scatter in IRX at a given $\beta$ (e.g.,
\citealt{reddy06a, reddy10, buat12, zeimann15, bouwens16b,
  shivaei20a}).

\begin{figure}
  \epsscale{1.1}
  \includegraphics[width=1.0\linewidth]{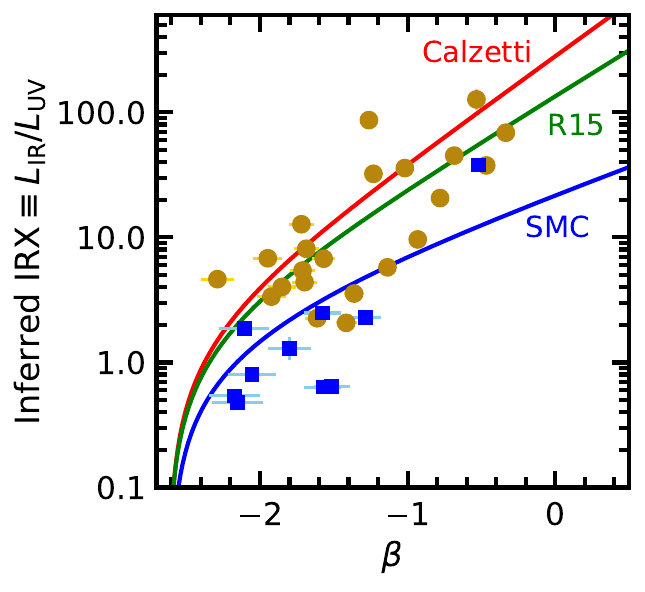}
    \caption{Relationship between inferred IRX and $\beta$, where the former is
      estimated from the dust-corrected nebular SFR, for galaxies
      within and outside the main sample, indicated by the gold circles and
      blue squares, respectively.  One of the galaxies in the main sample
      (COSMOS-4622) has $\beta = 1.63$ and ${\rm IRX} = 680$, and is
      not shown for clarity.  For the galaxies excluded from the main
      sample, $\sfrneb$ was computed assuming the Balmer decrement and
      the Galactic extinction curve (squares).  
The red, green, and blue lines indicate the
      predicted relationships between IRX and $\beta$ for the
      Calzetti, MOSDEF \citep{reddy15}, and SMC stellar attenuation curves,
      taken from \citet{reddy18a}.}
   \label{fig:irxbeta}
\end{figure}

\citet{reddy23a} demonstrated that the distribution of
nebular-inferred IRX and $\beta$ for CEERS galaxies at $z\sim 1-3$ is
generally consistent with the predictions of the SMC curve, generally
in line with direct measurements of IRX (e.g., \citealt{bouwens16a,
  reddy20, fudamoto20}; c.f., \citealt{mclure18, shivaei20b,
  bowler22}).  In \citet{reddy23a}, the nebular-inferred IRX was
determined by calculating the unobscured UV luminosity, $\luv$, and
then subtracting the implied unobscured UV SFR ($\sfruv$) from the
dust-corrected nebular-based SFR constrained from the Paschen lines.
This method provided an estimate of the dust-obscured SFR, which was
then converted to infrared luminosity, $\lir$, using the
\citet{kennicutt98} relation modified for a Chabrier IMF.  Finally,
the IRX ratio was derived by dividing the estimated $\lir$ by $\luv$.
The results of performing this same exercise for the AURORA sample are
shown in Figure~\ref{fig:irxbeta}.  The unobscured UV luminosity was
calculated directly from the best-fit SED at a rest-frame wavelength
of $1600$\,\AA\, and converted to an SFR assuming a luminosity-to-SFR
conversion factor that is specific to the adopted SPS model and the
age of the stellar population (e.g., \citealt{theios19}) as
constrained by the SED fitting.  This conversion factor varies from
$5.71\times 10^{-28}$\,$M_\odot$\,yr$^{-1}$\,erg$^{-1}$\,s\,Hz for the
youngest galaxy in the sample, GOODSN-17940, with an inferred age of
$\sim 10^{6.3}$\,yr, up to $5.95\times
10^{-29}$\,$M_\odot$\,yr$^{-1}$\,erg$^{-1}$\,s\,Hz for the oldest galaxies
in the sample, which have inferred ages of $\sim 10^{9.1}$\,yr.
Lastly, $\beta$ was calculated directly from the photometry of the
galaxies as described in \citet{reddy23c}.

There is one galaxy in the main sample of 24 for which $\sfruv$ is
comparable to $\sfrneb$, and for which $\lir$ could not be reliably
determined (GOODSN-25004).  The distribution of IRX and $\beta$ for
the remaining galaxies in the main sample---with the exception of
COSMOS-4622, which has $\beta = 1.63$ and ${\rm IRX} = 680$---is shown
in Figure~\ref{fig:irxbeta}, along with the predicted relations
between IRX and $\beta$ for the Calzetti, \citet{reddy15}, and SMC
curves, taken from \citet{reddy18a}.  The majority of galaxies fall in
between the IRX-$\beta$ predictions for the SMC and Calzetti curves,
with the \citet{reddy15} dust-curve prediction lying within this
range.  Some galaxies have IRX-$\beta$ that appear to favor an
SMC-like curve, while others favor the Calzetti curve.  

For the lower-SFR and less-dusty galaxies that fell out of the sample
(i.e., the excluded galaxies discussed in Section~\ref{sec:sfrcompare}
which do not have individual nebular attenuation curves), more than
half have $\sfrneb$ either comparable to or smaller than $\sfruv$.
The remainder are shown in Figure~\ref{fig:irxbeta}.  These galaxies
exhibit a distribution of IRX-$\beta$ that generally lie below both
the Calzetti and \citet{reddy15} predictions, and appear to be most
consistent with the SMC prediction.  This approach again suggests UV
dust corrections that are consistent with the predictions of the SMC
curve for the excluded galaxies.  While this IRX-$\beta$ analysis is
not independent of the SFR comparison discussed in
Section~\ref{sec:sfrcompare}, since $\sfrneb$ is used to estimate
$\lir$ and hence IRX, it does present an alternative way of
interpreting the data.

\begin{figure}
  \epsscale{1.1}
  \includegraphics[width=1.0\linewidth]{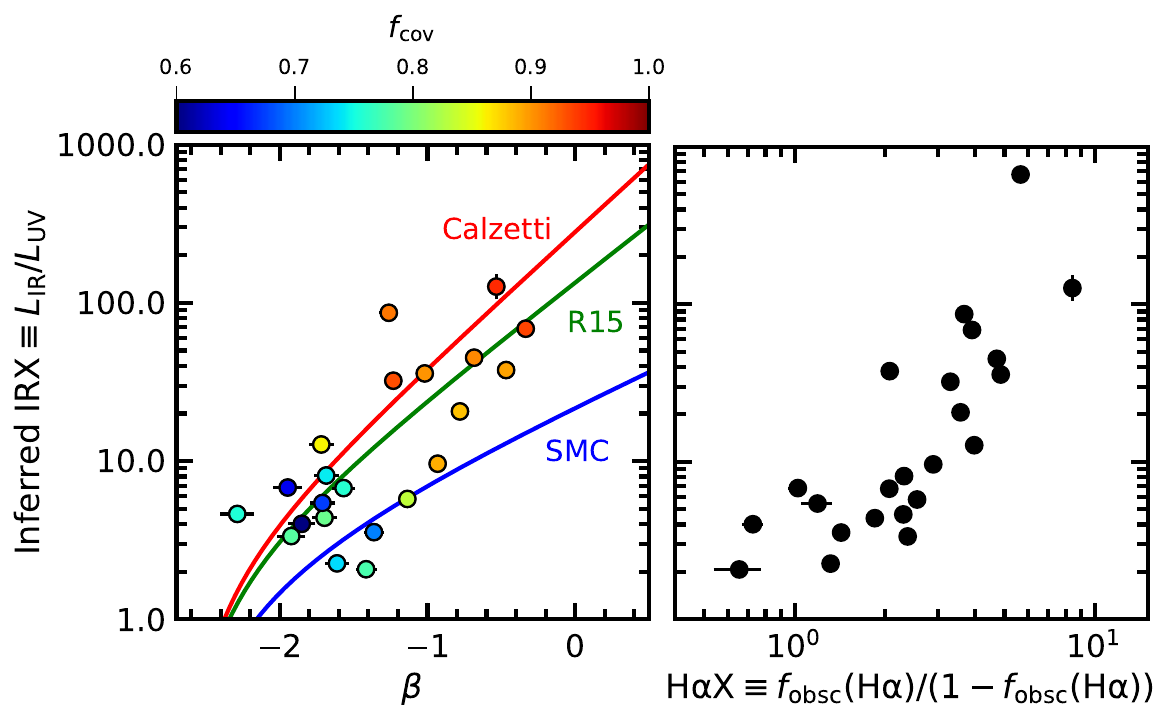}
    \caption{(Left:) Same as Figure~\ref{fig:irxbeta} where the main
      sample symbols have been color coded by $\fcov$.  (Right:) IRX
      versus HaX, where the latter is defined as the ratio of the
      dust-obscured $\ha$ luminosity to the observed $\ha$ luminosity
      (see text).}
   \label{fig:irxscatter}
\end{figure}

There are two additional points worth addressing.  First, galaxies
with the highest dust covering fractions ($\fcov$) appear to be
largely consistent with the Calzetti and MOSDEF \citep{reddy15} curves
(Figure~\ref{fig:irxscatter}).  Galaxies with lower dust covering
fractions, $\fcov \la 0.8$, are distributed evenly across the Calzetti,
MOSDEF, and SMC predictions.  In particular, we note that there are a
few galaxies with low $\fcov$ that appear to be more consistent with
SMC curve.  If the processes that lead to a non-unity covering
fraction of dust along the sightlines to OB associations also lead to
a non-unity covering fraction of dust towards the regions that
dominate the UV continuum emission, then these galaxies should follow
an attenuation curve that accounts for the presence of unreddened
sightlines.  At face value, this result appears to contradict the
interpretation of the SMC curve as an {\em extinction} curve, where
the dust is considered as a foreground screen with unity covering
fraction.  While the stellar reddening curve for these galaxies yields
a UV dust correction that is consistent with the SMC predictions, it
should not be viewed as a simple extinction curve.  Rather, it more
closely resembles an attenuation curve that accounts for the
contribution of unreddened sightlines within the galaxies.  SED
fitting that allows for both an unreddened and reddened component may
be used to investigate this issue further (see also \citealt{reddy16a,
  reddy16b, steidel18} for a similar treatment when modeling the
rest-frame UV spectra of high-redshift galaxies).

Second, just as IRX is the ratio of the dust-obscured UV luminosity
(or $\lir$) to the observed UV luminosity, we can define a similar
ratio for the $\ha$ luminosity.  Specifically, we define the quantity
$\ha$X as follows:
\begin{equation}
\ha{\rm X} \equiv \frac{f_{\rm obsc}(\ha)}{1-f_{\rm obsc}(\ha)},
\end{equation}
which is the ratio of the dust-obscured $\ha$ luminosity to the
observed $\ha$ luminosity.  As the UV luminosity has a significant
contribution from OB associations that dominate the nebular emission
(see Section~\ref{sec:differential}), IRX should correlate with
$\ha$X.  This expectation appears to be supported by the data, as
shown in the right panel of Figure~\ref{fig:irxscatter}.  While the
scatter in the correlation is artificially tight since both the
inferred IRX and $f_{\rm obsc}(\ha)$ depend on parameters constrained
from the covering-fraction model, the correlation itself indicates
that galaxies with higher $\ha$X also tend to have higher IRX.  Future
direct measurements of IR emission (e.g., from ALMA or similar
facilities) will enable a more detailed investigation of the strength
of this correlation.  If, for instance, there is a substantial scatter
or decoupling between IRX and $\ha$X, it could indicate differences in
the dust covering fractions between regions of star formation that
dominate the UV continuum (and IR) luminosity and those that dominate
the nebular emission.

\subsection{Impact on the Relationship between SFR and $M_\ast$}
\label{sec:sfrmstar}

\begin{figure*}
  \epsscale{1.1}
  \includegraphics[width=1.0\linewidth]{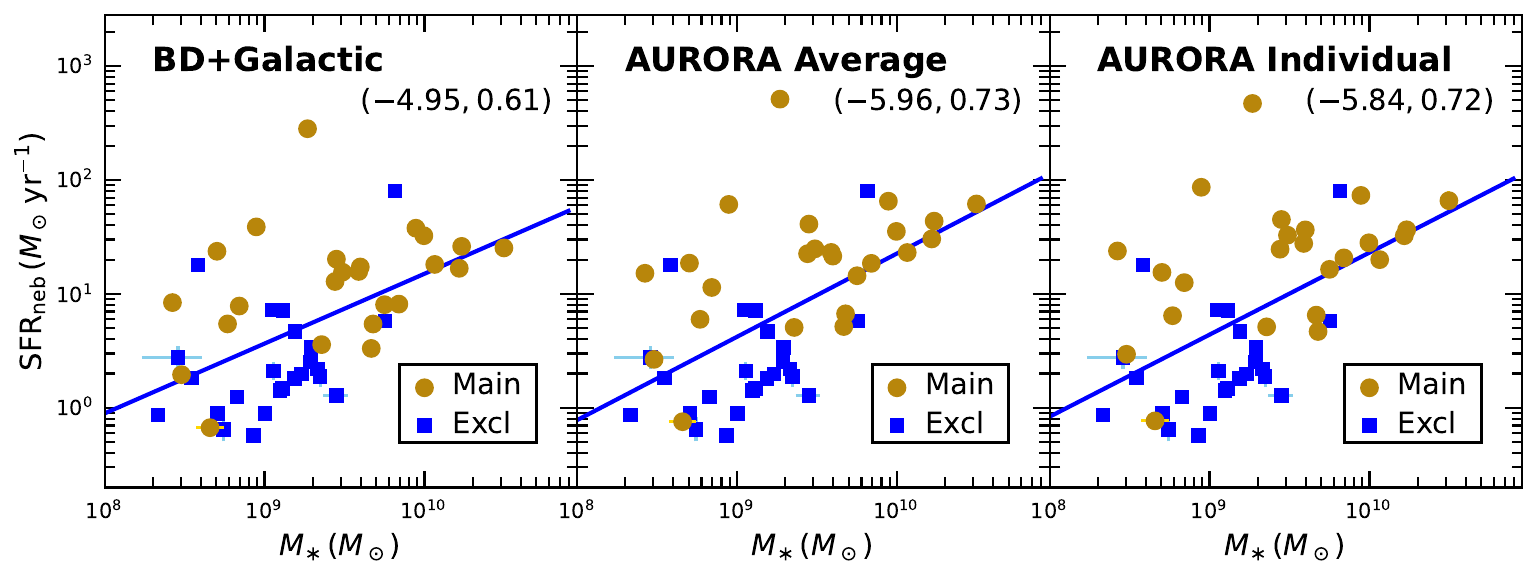}
    \caption{Relationship between SFR and $M_\ast$.  The 24 galaxies
      in the main sample are indicated by the circles in each panel,
      where $\sfrneb$ was estimated using the Balmer decrement and
      Galactic extinction curve (left panel), the average nebular dust
      attenuation curve found in this study (middle panel), and the
      individual nebular dust attenuation curves for the 24 galaxies
      (right panel).  Galaxies that were excluded from the main sample
      are indicated by the squares.  For these galaxies, $\sfrneb$ was
      estimated using the Balmer decrement and the Galactic extinction
      curve (shown as squares in all panels).  
The stellar masses for the main sample of 24 galaxies are obtained
from the SED fitting that assumes the MOSDEF stellar dust attenuation
curve, while those for the excluded sample assume the SMC extinction
curve, based on the results of Section~\ref{sec:sfrcompare}.  The blue
lines indicate the best fit relations to the combined sample of 24
galaxies and the excluded galaxies, with the intercepts and slopes of
the linear functions indicated as a pair in each panel.  The slope,
normalization, and scatter in the SFR-$M_\ast$ relation do not appear
to be sensitive to the methodology used to compute the dust-corrected
$\sfrneb$ for galaxies in the AURORA sample.}
   \label{fig:sfrmstar}
\end{figure*}

A key focus of galaxy evolution studies is the relationship between
SFR and stellar mass, $M_\ast$, as the functional form and scatter
of this relation can provide important constraints on the baryon cycle
(i.e., cold gas accretion and feedback) and the stochasticity of star
formation (e.g., \citealt{york00, brinchmann04, noeske07, dave08,
  reddy12b, speagle14, whitaker14, shivaei15b, santini17, atek22,
  clarke24, neufeld24}, among many others).  There are two aspects of
the so-called ``main sequence'' of star-forming galaxies that are
worth investigating in the context of the present analysis: the impact
of heavily-reddened star formation on the slope and scatter of the
SFR-$M_\ast$ relation, and the consequences of assuming individual
nebular dust attenuation curves on the scatter in the SFR-$M_\ast$
relation.

Figure~\ref{fig:sfrmstar} shows the relationship between SFR and
$M_\ast$ for both the main sample of 24 galaxies and the excluded
sample of 25 galaxies (Section~\ref{sec:sfrcompare}).  Based on the
results discussed in Section~\ref{sec:sfrcompare}, stellar masses for
the main sample were obtained from SED fitting that assumed the MOSDEF
stellar attenuation curve, as this curve provides the best agreement
between the nebular and SED-based SFRs.  Similarly, the stellar masses
for the excluded galaxies were obtained from SED fitting with the SMC
extinction curve.  The particular choice of stellar masses adopted for
the two subsamples does not affect the subsequent discussion.

For the excluded galaxies, the nebular SFRs were estimated using the
Balmer decrement and the Galactic extinction curve (squares).
In the left panel of Figure~\ref{fig:sfrmstar}, the SFRs for the main
sample of 24 galaxies are also estimated using this method.  In the
middle panel, the SFRs for the main sample are estimated using the
average nebular dust attenuation curve.  Finally, in the right panel,
the SFRs for the main sample are estimated using the individual
nebular attenuation curves for each galaxy.  As before, $\sfrneb$ for
galaxies in the main sample and the excluded sample were computed
based on the $N({\rm H})$ for the best-fitting SED model assuming the
MOSDEF attenuation curve and SMC extinction curve, respectively.

Relative to the traditional method of estimating $\sfrneb$ using the
Balmer decrement and the Galactic extinction curve, the assumption of
the average nebular attenuation curve derived in this study results in
systematically higher $\sfrneb$ due to the presence of highly-reddened
star formation, as discussed earlier.  We do not find a significant
trend between the difference in $\sfrneb$ calculated using the two
methodologies and the stellar mass among the galaxies in the main
sample.  Thus, the slope of the linear relation between $\sfrneb$ and
$M_\ast$ only changes by a small amount, increasing by $\approx 20\%$.
We do not find any
significant difference in the slope of the SFR-$M_\ast$ relation if
individual nebular attenuation curves are used for the dust
corrections rather than the average nebular attenuation curve.
Additionally, we do not find any significant difference in the scatter
in the SFR-M$_\ast$ relation if individual nebular attenuation curves
are used in lieu of the average nebular attenuation curve, consistent
with the conclusions of \citet{shivaei15b}.

The primary conclusion from this discussion is that---apart from a
modest increase in the slope of the SFR-$M_\ast$ relation when
assuming either the average nebular dust attenuation curve or the
individual nebular dust attenuation curves from AURORA---the scatter
in the relation appears to be largely unaffected by the methodology
used to correct for dust.  This conclusion holds true whether the
correction is made using the Balmer decrement and Galactic extinction
curve, the average nebular dust attenuation curve from AURORA, or the
individual nebular dust attenuation curves from AURORA.  Based on this
analysis, systematics in the assumed nebular dust attenuation curve
are a subdominant source of uncertainty in the slope and scatter of
the SFR-$M^\ast$ relation.

\subsection{Consequences for Calibrating JWST/MIRI-based SFRs}
\label{sec:miri}

One of the primary mission goals of JWST is to unveil dusty star
formation through observations at mid-IR wavelengths, made possible
with MIRI.  A key component of upcoming JWST/MIRI surveys is to
calibrate the relationship between mid-infrared emission and total SFR
(e.g., \citealt{alberts24}), either through complementary, direct
measurements of far-infrared and cool dust emission, or by measuring
the Paschen lines, which are less affected by dust obscuration
compared to the Balmer lines.  The large suite of Balmer and Paschen
lines available for the AURORA galaxies can be used to identify any
factors that may impact Paschen-line-inferred SFRs and, in turn, their use
as calibrators for mid-IR emission.

The joint analysis of the Balmer and Paschen lines for the AURORA
galaxies indicates that even these largely typical star-forming
galaxies may host a non-negligible fraction of heavily-reddened star
formation.  Dust corrections that assume the Balmer lines alone, along
with the Galactic extinction curve, systematically underpredict the
total SFR by $0.20$\,dex (Section~\ref{sec:effectonlineluminosities}).
Therefore, calibrating the relationship between mid-IR emission and
SFR using only the Balmer lines will likely result in mid-IR-inferred
SFRs that are systematically low.

While the Paschen lines are less affected by dust, our analysis shows
that the attenuation at the wavelength of the longest available
Paschen line for each galaxy (typically Pa5; i.e., Pa$\beta$) implies
a non-negligible upward correction of 0.11\,mag to account for dust
obscuration (Section~\ref{sec:effectonlineluminosities}).  Assuming
that these longer-wavelength lines are dust-free would not be accurate
for these galaxies.  In principle, multiple Paschen lines could be
used to compute a Paschen decrement, but this approach poses two
challenges.  First, similar to the Balmer decrement, one must still
assume a functional form for the nebular attenuation curve in order to
translate the Paschen decrement into a dust obscuration factor.
Second, since the Paschen lines are less sensitive to dust reddening,
by themselves they offer less constraining power on the reddening (see
R26a) and, consequently, the total dust obscuration.

Based on our analysis, detections of a large number of Balmer and
Paschen lines are needed to robustly constrain the effective
attenuation curve and/or covering-fraction model, and thus the
dust-corrected nebular SFRs.  Ideally, measurements of Pa4
(Pa$\alpha$), which are missing for all but one of the galaxies in our
sample, or, alternatively, independent measurements of the
dust-obscured luminosity, would be required to place more stringent
constraints on $\rv$.  

In summary, the presence of heavily-reddened star formation even in
typical star-forming galaxies at $z\sim 1-4$ implies that multiple
detections of Balmer and Paschen lines are crucial to reliably
constrain the dust-corrected nebular-based SFR, and thus use them to 
calibrate the relationship between mid-IR emission and SFR.
 Lastly, the $\hi$ recombination
lines may reflect a recent ($\la 10$\,Myr) upward or downward
variation in SFR that is not captured by the UV and IR emission.  This
effect may be relevant for bursty star-forming galaxies, and would
need to be accounted for
when establishing the relationship between mid-IR emission and total
SFR through the use of the $\hi$ recombination lines.

\section{Differential Reddening of the Stellar Continuum and Nebular Emission}
\label{sec:differential}

Considerable efforts have been made to understand the relationship
between the reddening of nebular lines and the stellar continuum in
galaxies (i.e., differential reddening), and its implications for the
spatial distribution of dust relative to stars of varying masses (see
Section~\ref{sec:intro} and references therein).  The traditional
interpretation of differential reddening suggests that the most
massive stars ($\ga 8$\,$M_\odot$), which power the nebular emission,
are short-lived and remain enshrouded by their parent GMCs, which
typically exhibit high dust column densities.  In contrast, the UV
continuum shortward of $\simeq 1800$\,\AA, in high-redshift, actively
star-forming galaxies arises from a broader range of stellar masses,
where older and less massive O, B, and massive A stars (spectral types
A0 and A1; e.g., \citealt{calzetti94, leitherer95}), have outlived the
GMCs from which they formed.  As a result, they are less affected by
reddening compared to the most massive O stars.  In this framework,
differential reddening can be linked to the strong dependence of the
stellar ionizing spectrum on stellar mass, and hence, main sequence
lifetime of stars, and how the latter compares with the GMC disruption
and/or cloud crossing timescale.

There are three points to consider regarding this ``timescale''
interpretation.  First, far-UV spectra of high-redshift star-forming
galaxies routinely show features associated with very massive O stars
(e.g., P-Cygni photospheric lines including $\civ$\,$\lambda\lambda
1548,1550$ and $\siiv$\,$\lambda\lambda 1393, 1402$) and Wolf Rayet
features (e.g., $\heii$\,$\lambda 1640$).  The presence of these
features implies that even OB associations containing the most massive
O stars (and presumably still embedded in their parent GMCs) must be
contributing significantly to the UV continuum---i.e., some fraction
of the youngest OB associations must lie along relatively
optically-thin sightlines for them to dominate the far-UV continuum.
Second, the implied non-negligible escape fractions of ionizing
photons from high-redshift galaxies (e.g., \citealt{steidel01,
  shapley06, shapley16, iwata09, vanzella10a, nestor11, nestor13,
  mostardi13, debarros16, steidel18, pahl21, marques24b}) also suggest
that some fraction of OB associations lie along sightlines with low
column densities of gas and likely dust.  Third, since the Balmer and
Paschen $\hi$ recombination lines are sensitive to the presence of the
same stars, differences in reddening inferred from these lines using
standard extinction curves (like the Galactic extinction curve,
Section~\ref{sec:intro}) imply spatial variations in the dust column
densities, or variations in dust optical depth, towards the youngest
OB associations (e.g., R26a, \citealt{lorenz25}).

Taken together, these observations indicate that, apart from the
timescale argument discussed above, differences in the reddening of
the UV stellar continuum and the nebular lines may be partly driven by
spatial variations in dust reddening towards the youngest (or
similarly-aged) OB associations.  Such variations may be due to a
nonuniform dust-to-gas ratio (or dust content) throughout the galaxy
and therefore the GMCs within a galaxy (e.g., \citealt{fetherolf20,
  gimenez22, fetherolf23, perezgonzalez23, cheng25, lorenz25,
  wozniak26}).  Spatial variations in the dust column density may also
arise if, for example, feedback from a supernova within one $\hii$
region affects the gas and dust covering fraction in a proximate or
overlapping, and slightly younger, $\hii$ region.  In this case, the
correlation between dust column density and the $\hii$ region (or
stellar) age will be weakened.

The multiple detections of Balmer and Paschen lines, and extensive
multi-wavelength rest-UV through rest-near-IR photometry available for
the AURORA sample, are used to investigate differential reddening
(Section~\ref{sec:ebmvcompare}).  The physical factors driving the
differences in the nebular and stellar dust attenuation curves is
discussed in Section~\ref{sec:rvcompare}.  Then, in
Section~\ref{sec:youngcurve}, we examine the relationship between the
nebular and stellar reddening curves, and implied color excesses, in
the simple case of a very young star-forming galaxy in the sample,
GOODSN-17940.

\subsection{Comparison of Nebular and Stellar Reddening}
\label{sec:ebmvcompare}

\begin{figure}
  \epsscale{1.2}
  \includegraphics[width=1.0\linewidth]{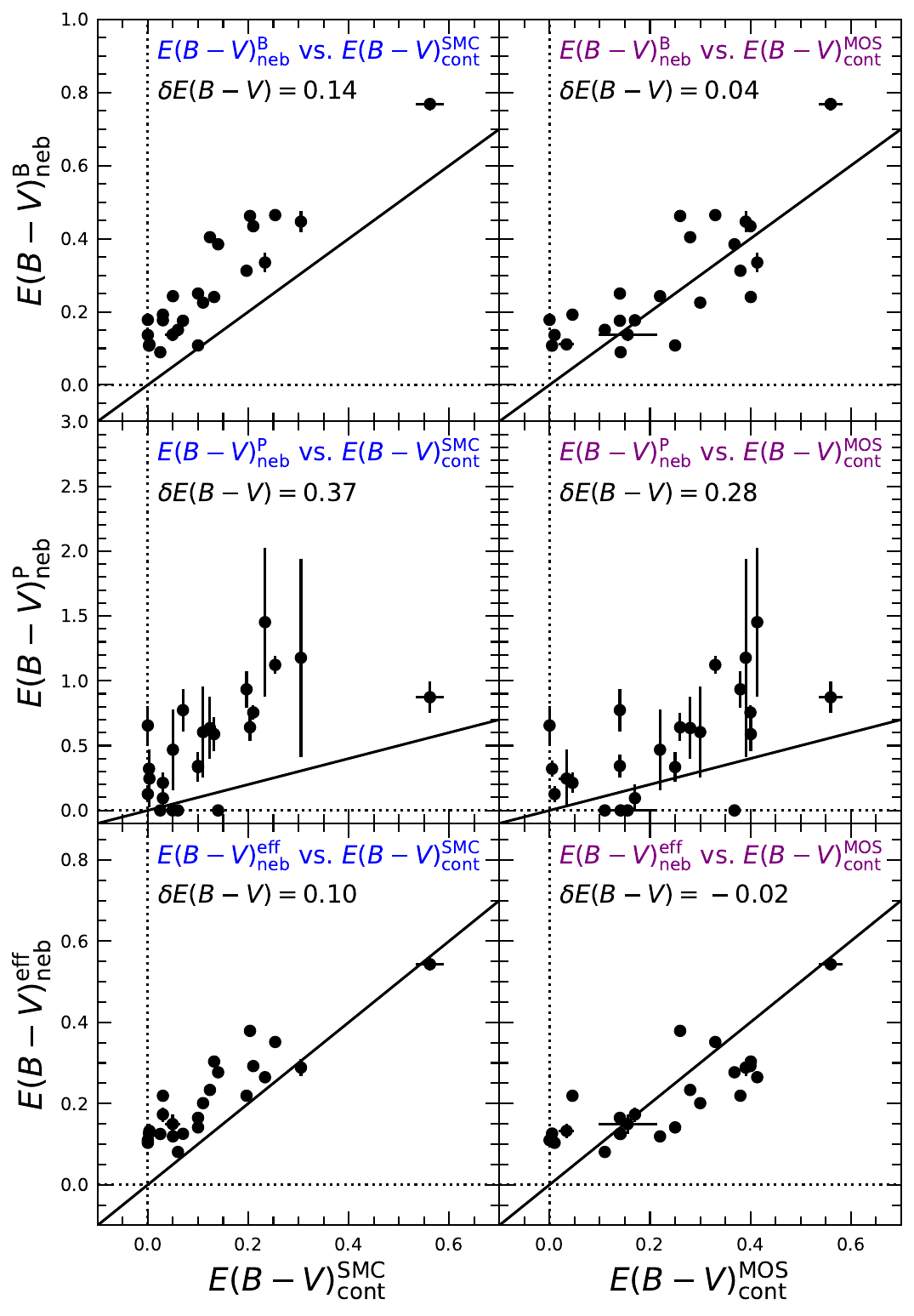}
    \caption{Comparison of $\ebmvneb$ and $\ebmvcont$.  The top,
      middle, and bottom rows show the results for $\ebmvneb$ computed
      with only the Balmer lines and the Galactic extinction curve,
      only the Paschen lines and the Galactic extinction curve, and
      all $\hi$ recombination lines and the individually-determined
      nebular dust attenuation curves, respectively.  The left and
      right columns show the results for $\ebmvcont$ returned from SED
      fitting assuming the SMC extinction curve and MOSDEF attenuation
      curve, respectively.  The solid line in each panel indicates the
      one-to-one relation.  The median difference in $\ebmvneb$ and
      $\ebmvcont$, $\delta\ebmv$, is also indicated in each panel.
      The best available constraints on the shapes of the nebular and
      stellar attenuation curves imply that the nebular emission
      exhibits the same level of reddening as the stellar continuum;
      i.e., $\ebmvneb \simeq \ebmvcont$.}
   \label{fig:delebmv}
\end{figure}

In the physical picture just put forth, both the UV reddening,
$\ebmvcont$, and the nebular reddening, $\ebmvneb$, are sensitive to
the distribution of dust column densities, or dust optical depths,
along the sightlines towards OB associations.  Consequently, the UV
continuum reddening and nebular reddening are expected to correlate.
Figure~\ref{fig:delebmv} displays the comparison between nebular
reddening and stellar reddening, assuming three ways of computing the
nebular reddening and two ways of computing the stellar reddening.
The nebular reddening was computed from only the Balmer lines assuming
the Galactic extinction curve (top row), only the Paschen lines
assuming the Galactic extinction curve (middle row), and all lines
assuming the individual nebular attenuation curves calculated for each
of the 24 galaxies in the sample.  The stellar reddening was obtained
from SED fitting where the SMC extinction curve (left column) and
MOSDEF attenuation curve (right column) were adopted.  Results for the
Calzetti curve are not shown as this curve returns SED-based SFRs that
are systematically higher than the nebular-based SFRs
(Section~\ref{sec:sfrcompare}).  Each panel indicates the median
difference between the nebular and stellar reddening, $\delta\ebmv
\equiv \ebmvneb - \ebmvcont$.

Spearman tests indicate strong correlations between the nebular and
stellar reddening, irrespective of how the two are computed.  The
probabilities of null correlations range from $p= 5\times 10^{-5}$ to
$p= 1\times 10^{-7}$ when the Balmer lines or all lines are used to
compute $\ebmvneb$.  When only the Paschen lines are used to estimate
$\ebmvneb$, the probabilities of null correlations rise to $p=5\times
10^{-4}$ to $6\times 10^{-3}$.  Although the Paschen lines alone
provide loose constraints on $\ebmvneb$, the latter still have a high
significance of correlation with respect to $\ebmvcont$.  In general,
the level of scatter between $\ebmvneb$ and $\ebmvcont$ in this
analysis is substantially smaller than what has been found in previous
studies (e.g., \citealt{reddy15, theios19, reddy20, shivaei20b}).  While part of this
difference may arise from sample selection effects---where the current
sample is biased toward slightly more luminous and redder galaxies
compared to the parent AURORA sample (see
Section~\ref{sec:data})---it is also likely that the large
scatter found in previous studies can be attributed to the lower-$S/N$
$\hi$ line measurements used in those works.

Aside from the strong correlation between $\ebmvneb$ and $\ebmvcont$,
Figure~\ref{fig:delebmv} also demonstrates the obvious point that the
magnitude of differential reddening depends on the dust attenuation
curve assumed for the stellar continuum and the nebular emission
lines.  Unsurprisingly, the assumption of the SMC curve results in
$\ebmvcont$ that are systematically smaller, and hence, $\delta \ebmv$
that are systematically higher, than those obtained with the MOSDEF
attenuation curve.  This result stems from the steeper wavelength
dependence of the SMC curve at UV wavelengths relative to the MOSDEF
attenuation curve \citep{reddy15, shivaei20a}.  The dependence of
$\delta \ebmv$ on the stellar reddening curve has been noted elsewhere
(e.g., \citealt{reddy15, theios19, reddy20, shivaei20a}).  Our
analysis shows that $\delta \ebmv$ is also sensitive to the $\hi$
recombination lines used to compute $\ebmvneb$.  For a given stellar
reddening curve (e.g., either SMC or the MOSDEF curves), $\delta
\ebmv$ is larger when only the Paschen lines and the Galactic
extinction curve are used to estimate $\ebmvneb$.

As noted in Section~\ref{sec:sfrcompare}, the MOSDEF attenuation curve
yields the best agreement between $\sfrsed$ and $\sfrneb$ for the 24
galaxies in the main sample.  Assuming this curve yields $\ebmvcont$
that are similar on average to $\ebmvneb^{\rm B}$---as shown in the
top right panel of Figure~\ref{fig:delebmv}---implying that
$\delta\ebmv \simeq 0$.  At face value, the similarity of the nebular
and stellar reddening suggests that the relatively unreddened, or
modestly reddened, OB associations which dominate the observed Balmer
recombination lines, may also contribute significantly to the UV
continuum.  In contrast, the reddening of the UV continuum is
systematically lower than the reddening deduced from the Paschen lines
(middle right panel of Figure~\ref{fig:delebmv}), similar to how the
reddening of the Balmer lines is systematically lower than the
reddening of the Paschen lines (Section~\ref{sec:intro} and R26a).  In
this case, the heavily reddened OB associations probed by the Paschen
lines contribute fractionally less light to the shorter wavelength
Balmer lines and, therefore, necessarily contribute fractionally less
light to the UV continuum.\footnote{The reddening calculated using the
  individual nebular attenuation curves is not sensitive to the lines
  (e.g., Balmer or Paschen) used in the calculation since the shapes
  of these curves account for optical-depth differences between the
  lines (see discussion in R26a).}
These results are consistent with a framework in which the differences
or similarities in the reddening of the continuum and nebular lines
are ostensibly driven by variations in the dust optical depths towards
OB associations.  

These results stand in contrast with many previous studies of both
local and high-redshift galaxies that have, in most cases, found the
nebular lines to be more heavily reddened that the stellar continuum
\citep{fanelli88, calzetti97, calzetti00, forster09, yoshikawa10,
  wild11, wuyts11, kashino13, kreckel13, price14, reddy15, debarros16,
  buat18, shivaei20b, reddy20, fetherolf21, lorenz23, lorenz24}.
Here, we have used the most accurate measurements of the shapes of the
individual nebular attenuation curves---which were not previously
obtainable---to deduce that the differential reddening between the
Balmer lines and the UV stellar continuum is negligible, at least for
the galaxies in our sample.

One very important and often overlooked caveat in this discussion is
the ambiguity of using differential reddening alone as an indicator of
the dust distribution in galaxies.  The reason for this ambiguity is
that the reddening depends on the shape of the attenuation curve,
which may differ between the nebular lines and the stellar continuum
(see Section~\ref{sec:rvcompare} for further discussion).  To more
accurately assess how nebular lines and the stellar continuum are
obscured by dust, it is best to consider the total attenuation,
$A(\lambda)$, at a fixed wavelength.  This approach was taken by
\citet{reddy15}, who directly compared the attenuation in magnitudes
of nebular $\ha$ emission with continuum photons at the same
wavelength for galaxies in the MOSDEF sample.  They found that the
attenuation of nebular emission exceeded that of the continuum on
average, with a strong dependence on the nebular SFR.  A similar
behavior is observed for the galaxies in the current sample, where the
attenuation of $\ha$ emission, $A_{\rm neb}^{\rm eff}(\ha)$, exceeds
that of the continuum at the same wavelength, $A_{\rm cont}(\lambda =
6564\,{\rm \AA})$, by $\simeq 0.85$\,mag on average, assuming the
MOSDEF attenuation curve for the reddening of the stellar continuum.

However, the challenge here is that less massive and older stars may
contribute significantly to the optical continuum photon flux at the
same wavelength as $\ha$.  As a result, the difference in attenuation
of nebular emission and the optical continuum may arise from the fact
that the former is powered by the youngest stars, while the latter may
have a significant contribution from older and less attenuated stars.
The ``differential attenuation'' calculated in the manner above will
not directly reveal how the UV continuum from OB associations is
reddened relative to the nebular emission.  Future work, using nebular
recombination emission lines at rest-frame UV wavelengths (e.g.,
$\heii\,\lambda 1640$ relative to $\heii\,\lambda 4686$;
\citealt{leitherer19}), will provide constraints on the rest-UV shape
of the nebular attenuation curve and offer a more direct comparison
between the reddening/attenuation of the nebular regions and the UV
continuum.

\subsection{What drives the differences between nebular and stellar dust attenuation curves?}
\label{sec:rvcompare}

In the previous section, we highlighted the ambiguity in interpreting
differential reddening.  The average offset between nebular and
stellar reddening is close to zero for galaxies in the AURORA sample,
suggesting that the UV continuum and nebular emission may originate
from similar spatial regions within these galaxies.  However, as
discussed in Section~\ref{sec:sfrcompare}, the stellar dust
attenuation curve that provides the best overall agreement between
SED-based (or UV-based) and nebular-based SFRs is the MOSDEF curve.  This
curve has an $\rv = 2.51$ \citep{reddy15}, which is significantly
lower than the range of $\rv\simeq 3.2-16.4$ inferred for nebular dust
attenuation curves (R26a).  Because $\rv$ is sensitive to the fraction
of unreddened emission contributing to the observed line or continuum
flux, this discrepancy in $\rv$ implies that the UV stellar continuum
and nebular emission are not fully spatially coincident, even though
their inferred reddening values are similar.

\begin{figure}
  \epsscale{1.0}
  \includegraphics[width=1.0\linewidth]{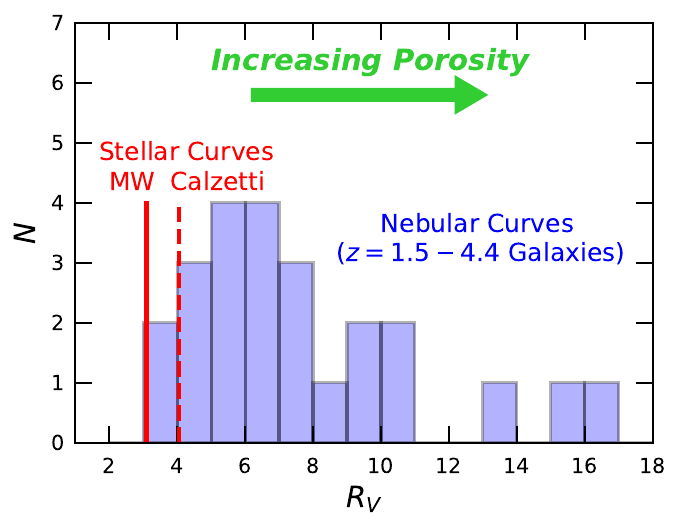}
    \caption{Comparison of the distribution of $\rv$ found for the
      nebular attenuation curves of the 24 galaxies in AURORA (blue
      histogram) and the $\rv$ of commonly adopted extinction and
      attenuation curves for the stellar continuum (red vertical
      lines).  The SMC extinction curve has $\rv=2.74$
      \citep{gordon03}, slightly less than that of the Milky Way ($\rv
      = 3.1$; \citealt{cardelli89}).  Increased ISM porosity towards
      massive-star sightlines would naturally explain the elevated
      $\rv$ of nebular attenuation curves relative to the $\rv$ of
      stellar dust attenuation curves.}
   \label{fig:rvcomphist_paper2}
\end{figure}

The pronounced difference between the $\rv$ values of nebular
attenuation curves and those of commonly adopted stellar attenuation
curves is illustrated in Figure~\ref{fig:rvcomphist_paper2}.  The
significantly higher $\rv$ inferred for nebular attenuation curves
suggests that sightlines towards OB associations contain a larger
fraction of unreddened light, indicative of a more porous dust and gas
distribution compared to that affecting the stellar continuum.  If
this porosity is driven by strong feedback from massive
stars---through processes such as photoionization, stellar winds, and
supernovae---it is perhaps not unexpected that sightlines towards
these massive stars (or $\hii$ regions) would probe a more porous ISM
than those associated with the broader stellar continuum.  This
connection between ISM porosity and $\rv$ is a natural explanation for
the high $\rv$ exhibited by nebular dust attenuation curves.  This
interpretation is consistent with previous studies showing that
greater ISM inhomogeneity leads to flatter attenuation curves (i.e.,
higher $\rv$; see \citealt{salim20} and references therein).

\subsection{Testing the Convergence of the Nebular and Stellar Attenuation Curve
for Young Galaxies}
\label{sec:youngcurve}

In the previous section, we discussed how differential reddening may
arise from differences in the main-sequence lifetimes of stars
dominating the nebular and UV continuum emission, and how those
lifetimes compare to GMC cloud disruption and/or cloud crossing
timescales.  We also explored how differential reddening could result
from variations in dust optical depth towards the youngest OB
associations.  Finally, we highlighted the challenge of interpreting
differential reddening and attenuation.  In principle, these
complications can be circumvented when considering very young
star-forming galaxies, where the light from OB associations dominates
the stellar continuum at all wavelengths.  For such galaxies, a simple
expectation is that the shape of the effective attenuation curve for
the stellar continuum---and the reddening implied by that
curve---should match those of the ionized regions.

This expectation can be tested directly for the youngest galaxy in the
sample, GOODSN-17940, at $z=4.411$.  This galaxy has an SED-derived
age ranging from 2 to 6\,Myr, depending on the assumed stellar
reddening curve (i.e., the SMC, MOSDEF, or Calzetti curves).
Additionally, as pointed out by \citet{sanders25}, the very high
rest-frame equivalent widths of $\ha$ and $\hb$ ($W_{\lambda} \simeq
1600$ and $280$\,\AA, respectively) imply extremely young ages of $\la
6$\,Myr.  Furthermore, the continuum emission from the galaxy is
detected with high S/N in the NIRSpec spectrum, and shows evidence of
the Balmer jump, a feature commonly observed in young galaxies with a
strong nebular continuum.  Thus, the SED fit, equivalent widths of
$\ha$ and $\hb$, and the nebular continuum shape all ostensibly
suggest a young age of $<10$\,Myr.

\citet{sanders25} present the nebular dust attenuation curve for
GOODSN-17940.  They assumed that the reddening of the stellar continuum
matches that of the nebular lines in order to constrain the shape of
the nebular reddening curve at UV wavelengths.  Here, we take
advantage of the high-S/N continuum spectrum of this galaxy to
independently compute the shape of the stellar dust attenuation curve
over the full optical through near-IR wavelength range and compare it
to the nebular attenuation curve.

Just as the derivation of the nebular attenuation curve relies on
knowledge of the intrinsic $\hi$ recombination emission line ratios,
the derivation of the stellar attenuation curve depends on knowledge
of the intrinsic spectrum of the galaxy, or the assumption that this
intrinsic spectrum is similar to that of galaxies with different
amounts of reddening (e.g., \citealt{noll09, kriek13, scoville15,
  reddy15, zeimann15, shivaei20a}).  The intrinsic spectrum is
influenced by several factors, most notably the star-formation history
of the galaxy, which may be complex and difficult to constrain.
However, these complexities in the star-formation history are not a
concern for a galaxy as young as GOODSN-17940, where the limited
inferred age of the galaxy makes the intrinsic spectrum relatively
straightforward to model.

The intrinsic spectrum of GOODSN-17940 was modeled using the BPASS
constant star-formation model which incorporates binary stellar
evolution, with an age of $3$\,Myr, a stellar metallicity of $Z_\ast =
0.001$ (i.e., $0.07Z_\odot$)\footnote{This choice of metallicity is
  motivated by previous studies that suggest significantly subsolar
  stellar metallicities that lag the oxygen abundances due to
  $\alpha$-enhanced abundance patterns inferred for high-redshift
  star-forming galaxies (e.g., \citealt{steidel16, cullen19,
    topping20a, topping20b, reddy22}).}, and a $100$\,$M_\odot$ upper
mass limit of the IMF (see below for a discussion of how varying these
assumptions influences the analysis).  The predicted intrinsic nebular
continuum was generated by using the SPS model as input for Cloudy
v23.01 \citep{ferland17, gunasekera23} and adopting a gas-phase oxygen
abundance of $Z_{\rm neb} = 0.4Z_\odot$ and an ionization parameter of
$\log U = -2.5$, consistent with typical values found in galaxies at
$z\sim 2-4$ (e.g., \citealt{steidel16, sanders16a, strom17, strom18,
  topping20a, topping20b, jeong20, reddy22, shapley24}).

The calculation of the stellar reddening curve proceeded in two ways.
In the first, the intrinsic nebular continuum shape predicted by
Cloudy was scaled according to the ionizing photon rate, $Q({\rm H})$,
inferred from the dust-corrected $\hi$ recombination lines.  This
scaled nebular continuum spectrum was then attenuated by dust based on
the effective nebular attenuation curve of GOODSN-17940 and the
corresponding $\ebmvneb^{\rm eff}$.  The reddened nebular continuum
was subtracted from the NIRSpec spectrum to yield a
stellar-continuum-only spectrum.  The latter spectrum (denoted by
$f(\lambda)$) and the intrinsic SPS model (denoted by $f_0(\lambda)$)
were then combined using the following equation to compute the
relative magnitude of attenuation of the continuum as a function of
wavelength:
\begin{eqnarray}
A_{\rm cont}'(\lambda_2) & \equiv & 2.5\left[\log_{10}\left(\frac{f(\lambda_1)}{f(\lambda_2)}\right) - \log_{10}\left(\frac{f_0(\lambda_1)}{f_0(\lambda_2)}\right)\right] + 1,
\label{eq:amag4}
\end{eqnarray}
where $\lambda_1$ is the reference wavelength, taken to be the wavelength of $\ha$.  
A fifth-order polynomial was fit to
$A_{\rm cont}'(\lambda)$ to infer $\ebmvcont$ and the effective attenuation curve of the continuum, $k_{\rm
  cont}'(\lambda)$, using the following:
\begin{eqnarray}
k_{\rm cont}'(\lambda) & = &\frac{A_{\rm cont}'(\lambda)}{A_{\rm cont}'(B) - A_{\rm cont}'(V)},
\label{eq:kprime4}
\end{eqnarray}
where we have set $\lambda = \lambda_2$ and used the definition that
$\ebmvcont = A_{\rm cont}'(B) - A_{\rm cont}'(V)$.  These equations
are analogous to Equations 8 and 9 in R26a.

In the second method, the intrinsic nebular continuum spectrum
predicted by Cloudy was directly added to the SPS model. Adding the
nebular continuum to the SPS model results in a slightly redder
intrinsic UV slope.  The resulting spectrum and the unmodified NIRSpec
spectrum were combined using Equation~\ref{eq:amag4} to compute
$A_{\rm cont}'(\lambda)$.  The computation of $k_{\rm cont}'(\lambda)$
proceeded in the same way as the first method.  Both approaches yield
$k_{\rm cont}'(\lambda)$ that are identical within the uncertainties.
The measurement uncertainties were determined by perturbing the
NIRSpec spectrum according to the error spectrum many times, and
recomputing $k_{\rm cont}'(\lambda)$ for each of these realizations.

\begin{figure}
  \epsscale{1.2}
  \includegraphics[width=1.0\linewidth]{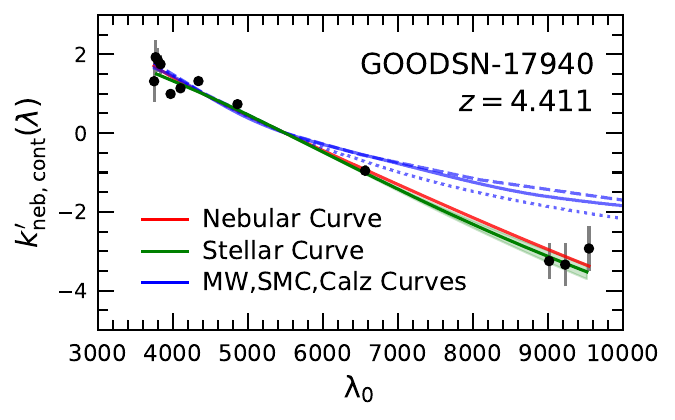}
    \caption{Comparison of $k_{\rm neb}'(\lambda)$ (red curve) and
      $k_{\rm cont}'(\lambda)$ (green curve) for GOODSN-17940 at
      $z=4.411$.  The $1\sigma$ uncertainty in the latter is indicated
      by the green shaded region.  For clarity, the uncertainty in
      $k_{\rm neb}'(\lambda)$ is not shown.  Individual $k_{\rm
        neb}'(\lambda)$ points are denoted by the black symbols.  Also
      shown are the Galactic, SMC, and Calzetti curves (solid, dotted,
      and dashed blue lines, respectively), shifted so that they pass
      through zero at $5500$\,\AA. The nebular and stellar dust
      attenuation curves for GOODSN-17940 are essentially identical
      over the wavelength range $3600\la \lambda_0/{\rm \AA} \la
      9600$, and depart from the behavior of other common extinction
      and attenuation curves.}
   \label{fig:gn17940curves}
\end{figure}

Figure~\ref{fig:gn17940curves} compares the relative attenuation
curves of the nebular lines and the stellar continuum of GOODSN-17940.
The two are identical within the uncertainties over the entire optical
through near-IR wavelength range.  Most notably, both the nebular and
stellar reddening curves exhibit the same significant departure from
the shapes of other standard extinction and attenuation curves (e.g.,
Galactic, SMC, and Calzetti curves) at $\lambda_0 \ga 6000$\,\AA.
Moreover, the inferred reddening of the stellar continuum, $\ebmvcont
= 0.31\pm 0.02$, is similar to the effective nebular reddening,
$\ebmvneb^{\rm eff} = 0.28\pm 0.01$.  This analysis indicates that the
nebular line emission and stellar continuum are both subject to the
same reddening curve and exhibit the same color excess, strongly
suggesting that the stellar continuum arises from the same OB
associations that power the nebular line emission, in accordance
with the expectations for this very young galaxy.

Here, we discuss the impact of varying the assumptions used to compute
the intrinsic spectrum and nebular continuum.  In general, the
strength of the Balmer jump in the observed spectrum is only matched
when considering models with ages $\la 10$\,Myr.  Furthermore, varying
$Z_{\rm neb}$ or $U$ does not significantly alter either the shape of
$k_{\rm cont}'(\lambda)$ or $\ebmvcont$.  In particular, the intensity
of the nebular continuum is determined by the ionizing photon rate and
is therefore insensitive to variations in $Z_{\rm neb}$ or $U$.
Adopting a higher stellar metallicity, such as solar metallicity
($Z_\ast = 0.02$), for the SPS model results in $k_{\rm
  cont}'(\lambda)$ similar in shape to the one derived using the
default assumptions described above, but with a higher continuum
reddening of $\ebmvcont = 0.40 \pm 0.02$.  At higher stellar
metallicities, the nebular continuum makes a smaller fractional
contribution relative to the stellar continuum due to a softer
ionizing spectrum.  However, we consider such a high stellar
metallicity to be unrealistic for this galaxy given its apparent very
young age and strong nebular continuum emission.

Using a model with an IMF upper-mass cutoff of $300$\,$M_\odot$ yields
$k_{\rm cont}'(\lambda)$ that is statistically consistent with the one
obtained with an upper-mass cutoff of $100$\,$M_\odot$, but with a
slightly bluer $\ebmvcont = 0.25\pm 0.02$.  In this case, a higher
upper-mass cutoff results in a stronger nebular continuum and redder
intrinsic UV slope.  Thus, less reddening is required to reproduce the
observed spectrum.  The BPASS single stellar population
models yield a continuum reddening that is $\simeq 0.05$\,mag bluer compared
to that obtained from the binary stellar evolution models.

In summary, altering the initial assumptions for the intrinsic SPS
model and the nebular continuum emission does not affect
the shape of $k_{\rm cont}'(\lambda)$.  In all cases, this curve
exhibits a strong departure from standard extinction and
attenuation curves at $\lambda \ga 6000$\,\AA, and matches the nebular
dust attenuation curve derived from all the available Balmer and
Paschen lines for GOODSN-17940.

This analysis points to nebular and stellar attenuation curves that
are essentially identical for the one very young galaxy in the sample.
However, it remains uncertain whether this conclusion holds for young
galaxies in general.  For instance, such galaxies could still exhibit
differences between the nebular and stellar attenuation curves
depending on the dust mixing timescale in the ISM.  Emission from the
ionized gas, including the nebular recombination lines, arises from
spatially extended regions with a volume filling fraction many orders
of magnitude larger than that of the stars.  If the dust mixing
timescale on the physical scale of individual or overlapping $\hii$
regions is comparable to or longer than a few Myr, the dust
distribution is likely to vary across these spatial scales.  This
effect could lead to variations in dust optical depth from the centers
to the outskirts of $\hii$-region complexes.  Given the complex
structure of $\hii$ regions and the potential for overlapping $\hii$
regions that span large spatial extents, long dust mixing timescales
may cause dust optical depths to vary on small (pc) spatial scales.
Therefore, we cannot rule out the possibility of discrepancies between
the nebular and stellar reddening curves, even for the youngest
galaxies at high redshift.

Finally, we should point out that the ostensibly young age of
GOODSN-17940 appears to be at odds with its high dust reddening
($\ebmvneb \simeq 0.28$), moderate gas-phase oxygen abundance
($12+\log({\rm O/H}) \simeq 8.3$, $\simeq 0.4Z_\odot$), and moderate
stellar mass ($M^\ast \simeq 10^9$\,$M_\odot$), values that are
similar to those of older galaxies in the AURORA sample --- i.e., one
may have expected little dust reddening and a low O abundance for such
a ``young'' galaxy.  While the concordance between the stellar and
nebular attenuation curves of GOODSN-17940 implies that the
dust-mixing timescale on the scale of $\hii$ regions is very short
(less than the age of the galaxy, $\la 6$\,Myr), a larger sample of
young galaxies is needed to explore this scenario further.

\section{Conclusions and Recommendations}
\label{sec:conclusions}

In R26a, we used deep JWST/NIRSpec
spectra from the AURORA survey to measure multiple $\hi$ Balmer and
Paschen lines for 24 individual $z=1.4-4.4$ galaxies, compute their
effective nebular attenuation curves, and place constraints on the
dust covering fraction towards the ionized regions in these galaxies.
In this second paper, we examine the implications of the nebular dust
attenuation curves for dust-corrected line luminosities and ratios,
total nebular-based SFRs, and differential reddening of the nebular
lines and stellar continuum.  We also perform a joint analysis of the
nebular and stellar attenuation curves for the youngest galaxy in the
sample, GOODSN-17940.  The primary results of this analysis are as
follows:

\begin{itemize}

\item The individual nebular dust attenuation curves yield effective
  reddening, $\ebmvneb^{\rm eff}$, that are systematically lower---and
  total attenuation that is systematically higher---than the values
  obtained by combining the Balmer decrement with the commonly assumed
  Galactic extinction curve (Section~\ref{sec:bdcompare} and
  Figure~\ref{fig:bdcompare}).  As a consequence, dust-corrected $\ha$
  luminosities, and hence the total ionizing photon rates, $Q({\rm
    H})$, are $\approx 0.20$\,dex larger when assuming the individual
  nebular dust attenuation curves than when interpreting the Balmer
  decrement using the Galactic extinction curve
  (Section~\ref{sec:effectonlineluminosities}).  This offset is due to
  the presence of dusty star formation that can be reliably probed via
  the $\hi$ Paschen lines.

\item The dust-corrected line ratios of O32, R23, and S32, which are
  commonly used to infer ionization parameter and oxygen abundance,
  are offset by $\la 0.05$\,dex compared to the values obtained with
  the Balmer decrement and Galactic extinction curve
  (Section~\ref{sec:effectonlineratios} and Appendix~\ref{sec:app1}).
  Applying the individual nebular attenuation curves yields an
  auroral-to-strong-line ratio of $\oiii\lambda 4364/\oiii\lambda
  5008$ that is on average 0.03\,dex lower than that inferred using
  the Balmer decrement and Galactic extinction curve to correct for
  dust.  On the other hand, the auroral-to-strong line ratio of
  $\oii\lambda\lambda 7320, 7330/\oii\lambda\lambda 3727,3730$ is
  approximately 0.07\,dex higher on average, and can be as high as
  0.19\,dex for individual objects, when assuming the dust corrections
  implied by the individual nebular attenuation curves relative to
  that obtained with the Balmer decrement and Galactic extinction
  curve.  However, we find that the systematic bias in the derived
  oxygen abundance is $\la 0.15$\,dex, and is thus a subdominant
  source of uncertainty in oxygen abundance measurements for the
  typical galaxy in the AURORA sample.

\item The dust-corrected nebular-line-based SFRs ($\sfrneb$) exhibit
  positive correlations with $\fcov$, $\ebmvneb^{\rm cov}$, and the
  fraction of $\ha$ luminosity that is obscured by dust, $f_{\rm
    obsc}(\ha)/f_0(\ha)$ (Appendix~\ref{sec:app2} and
  Figure~\ref{fig:sfrtrends}).  For the 24 galaxies for which
  individual nebular attenuation curves were derived, the stellar
  reddening curve calculated from the MOSDEF sample \citep{reddy15,
    shivaei20b} provides the best average agreement between SED-based
  SFR ($\sfrsed$) and $\sfrneb$ (Section~\ref{sec:sfrcompare} and
  Figure~\ref{fig:sfrcompare}).  For the less-reddened and
  less-luminous AURORA galaxies that do not have individual
  nebular-attenuation-curve measurements, an SMC-like extinction curve
  for the stellar continuum yields the best average agreement between
  $\sfrsed$ and $\sfrneb$, assuming significantly sub-solar stellar
  metallicities.  Similar conclusions are reached when examining the
  inferred ratio of the dust-obscured UV (or IR) luminosity to the
  unobscured UV luminosity, ``IRX'' (Section~\ref{sec:irxbeta} and
  Figure~\ref{fig:irxbeta}).  Furthermore, our results imply that
  galaxies with higher IRX have larger dust covering fractions
  (Figure~\ref{fig:irxscatter}).

\item We find that the newly-derived nebular SFRs do not appreciably
  affect the slope and scatter of the relationship between SFR and
  $M_\ast$ compared to the values obtained with the typically assumed
  combination of the Balmer decrement and Galactic extinction curve
  (Section~\ref{sec:sfrmstar} and Figure~\ref{fig:sfrmstar}).

\item Though the galaxies in our sample are not exceptionally dusty,
  and are representative of typical ``main-sequence'' star-forming
  galaxies at $z\sim 1.5-4.4$---with $\log[M^\ast/M_\odot] = 8.5-10.5$
  and SFRs $\simeq 1$ to $400$\,$M_\odot$\,yr$^{-1}$---the longest
  wavelength Paschen line available for most of the galaxies (i.e.,
  Pa$\beta$) still suffers a non-negligible attenuation of $0.11$\,mag
  on average (Sections~\ref{sec:bdcompare} and
  \ref{sec:effectonlineluminosities}).  Thus, it would be incorrect to
  assume that the Paschen lines are not significantly attenuated by
  dust even for these modestly dusty galaxies.

\item The inclusion of the $\hi$ Paschen lines is important for
  probing optically thick star formation and determining total SFRs.
  As such, these lines should be considered when calibrating the
  relationship between mid-IR luminosity and total SFR
  (Section~\ref{sec:miri}).

\item The difference in reddening of the nebular lines and the stellar
  continuum (i.e., differential reddening) is on average negligible if
  one uses the individual nebular attenuation curves to derive
  $\ebmvneb^{\rm eff}$ and the MOSDEF stellar attenuation curve to
  derived $\ebmvcont$ (Section~\ref{sec:ebmvcompare} and
  Figure~\ref{fig:delebmv}).  We find that spatial variations in the
  dust optical depths towards similarly-aged OB associations may play
  a significant role in causing differential reddening, alongside
  timescale effects.

\item We highlight the stark contrast between the low $\rv\simeq 2-5$
  (or steepness) of commonly assumed stellar dust attenuation curves
  and the range of high $\rv\simeq 3-16$ (or flatness) of nebular dust
  attenuation curves.  The high $\rv$ of the latter implies a more
  porous dust and gas distribution towards OB sightlines---potentially
  established by strong feedback from massive stars---relative to the
  sightlines towards the broader stellar continuum
  (Section~\ref{sec:rvcompare}).

\item For the youngest galaxy in the sample, GOODSN-17940, at
  $z=4.411$, the inferred stellar reddening curve, and the associated
  $\ebmvcont$, are essentially identical to the effective nebular
  attenuation curve, and the associated $\ebmvneb^{\rm eff}$,
  respectively (Section~\ref{sec:youngcurve} and
  Figure~\ref{fig:gn17940curves}).  These results indicate that the
  stellar continuum suffers the same magnitude of attenuation as the
  nebular emission at any given wavelength, and that the OB
  associations powering the line emission also dominate the stellar
  continuum, in accordance with our expectations for this young
  galaxy.

\end{itemize}

{\em Based on this analysis, we recommend the following.  When
  multiple $\hi$ Balmer and Paschen emission lines are available, the
  nebular attenuation curve should be computed directly---e.g., using
  the methodology described in R26a (see also \citealt{reddy20}).
  This empirically derived curve can then be used to calculate the
  nebular reddening and dust correct nebular lines and line ratios in
  a manner that is self-consistent with the constraints from all the
  Balmer and Paschen lines.  If there are insufficient high-S/N $\hi$
  line detections to robustly determine the nebular attenuation
  curve---i.e., if the combination of several Balmer lines and at
  least one Paschen line is lacking---we recommend using the average
  nebular dust attenuation curve given by Equation~\ref{eq:avecurve}:
\begin{eqnarray}
\langle k_{\rm neb}(\lambda)\rangle & = & -7.241 +
\frac{17.002}{\lambda/\mu{\rm m}} - \frac{8.086}{(\lambda/\mu{\rm
    m})^2} \nonumber \\ 
& & + \frac{2.177}{(\lambda/\mu{\rm m})^3} -
\frac{0.319}{(\lambda/\mu{\rm m})^4} + \frac{0.021}{(\lambda/\mu{\rm
    m})^5}, \nonumber
\end{eqnarray}
valid over the wavelength range $0.35\la \lambda \la 1.28$\,$\mu$m.}

This analysis has focused on a subset of galaxies in the AURORA sample
for which robust constraints on the nebular attenuation curves could
be obtained.  More generally, the construction of composite spectra
across wider dynamic ranges in galaxy properties will allow us to
better understand the factors that influence the attenuation curves,
including possible variations in the dust grain composition and/or
size distribution.  Targeting samples at slightly lower redshifts
($z<1.67$) with coverage of Pa$\alpha$, which is measured for only one
of the 24 galaxies in this analysis, will enable more precise
measurements of $\rv$.

We have also highlighted the connection between the $\hi$
recombination line ratios and dust covering fraction and total
obscuration.  Pairing this analysis with direct dust (IR) luminosity
measurements will allow us to quantify the connection between UV- and
H$\alpha$ inferred dust obscuration fractions (e.g.,
Section~\ref{sec:irxbeta} and Figure~\ref{fig:irxscatter}), and
perform direct comparisons of covering fractions between regions
dominating the nebular and UV continuum emission, lending further
insight into the distribution of dust in galaxies.  Analyses of larger
samples of young galaxies will reveal whether the similarity in the
nebular and stellar reddening curves of GOODSN-17940 extends to other
young galaxies where OB associations are expected to dominate the
stellar continuum.  These investigations will elucidate the processes
of dust and metal mixing on the spatial scales of $\hii$ regions.

\begin{acknowledgements}

This work is based on observations made with the NASA/ ESA/CSA James
Webb Space Telescope. The data were obtained from the Mikulski Archive
for Space Telescopes at the Space Telescope Science Institute, which
is operated by the Association of Universities for Research in
Astronomy, Inc., under NASA contract NAS5-03127 for JWST. The specific
observations analyzed can be accessed via doi:10.17909/hvne7139. We
also acknowledge support from NASA grant No. JWST-GO-01914.  Some of
the data products used in this analysis were retrieved from the Dawn
JWST Archive (DJA). DJA is an initiative of the Cosmic Dawn Center
(DAWN), which is funded by the Danish National Research Foundation
under grant DNRF140.

\end{acknowledgements}

\facility{{\em JWST}/NIRSpec}



\appendix

\section{Effect of Nebular Dust Attenuation Curves on Line Luminosities and Line Ratios}
\label{sec:app1}

\begin{figure*}
  \epsscale{1.20}
  \includegraphics[width=1.0\linewidth]{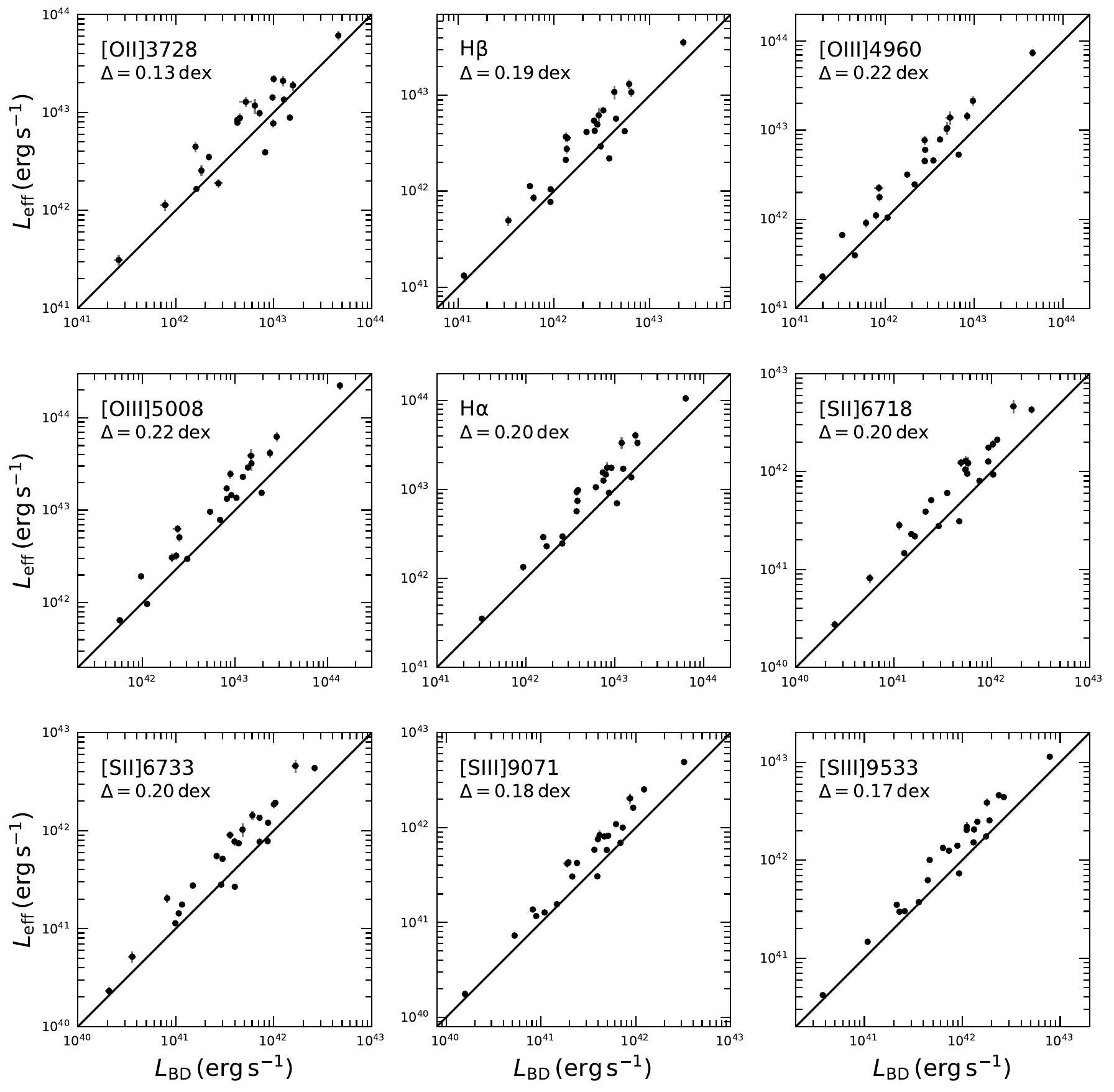}
    \caption{Comparison of dust-corrected line luminosities assuming
      the individual effective nebular attenuation curves ($L_{\rm
        eff}$) and the Balmer decrement and the Galactic extinction
      curve ($L_{\rm BD}$) for the following lines:
      $\oii\lambda\lambda 3727,3730$, $\hb$, $\oiii\lambda\lambda
      4960,5008$, $\ha$, $\sii\lambda\lambda 6718, 6733$, and
      $\siii\lambda\lambda 9071, 9533$.  The line of equality is
      indicated in black.  Each panel indicates the mean offset in
      luminosity ($\log L_{\rm eff} - \log L_{\rm BD}$) in dex.}
   \label{fig:lumcompare}
\end{figure*}

Figure~\ref{fig:lumcompare} compares the dust-corrected line
luminosities assuming the individual effective nebular attenuation
curves ($L_{\rm eff}$) and the Balmer decrement and the Galactic
extinction curve ($L_{\rm BD}$) for a number of important rest-frame
optical emission lines.  As noted in
Section~\ref{sec:effectonlineluminosities}, the values of $\log L_{\rm
  eff}$ are 0.13 to 0.22\,dex larger than $\log L_{\rm BD}$.

Note that the magnitude of the mean offset will depend on the
reddening distribution of galaxies in the sample.  Since the sample
was constructed to include galaxies where there is non-negligible
reddening implied by the $\hi$ recombination lines (in order to ensure
robust constraints on the nebular attenuation curve;
Section~\ref{sec:data}), the mean offset will be larger than it would
have been had there been no restriction on the reddening of the
galaxies.  For example, if we consider the larger subset of 40
galaxies in the AURORA sample with $>3\sigma$ detections of both $\ha$
and $\hb$ and at least one detected Paschen line, the mean offsets in
the dust-corrected $\ha$ and Paschen-line luminosities are lower at
$0.16$ and $0.07$\,dex, respectively.

\begin{figure*}
  \epsscale{1.00}
  \includegraphics[width=1.0\linewidth]{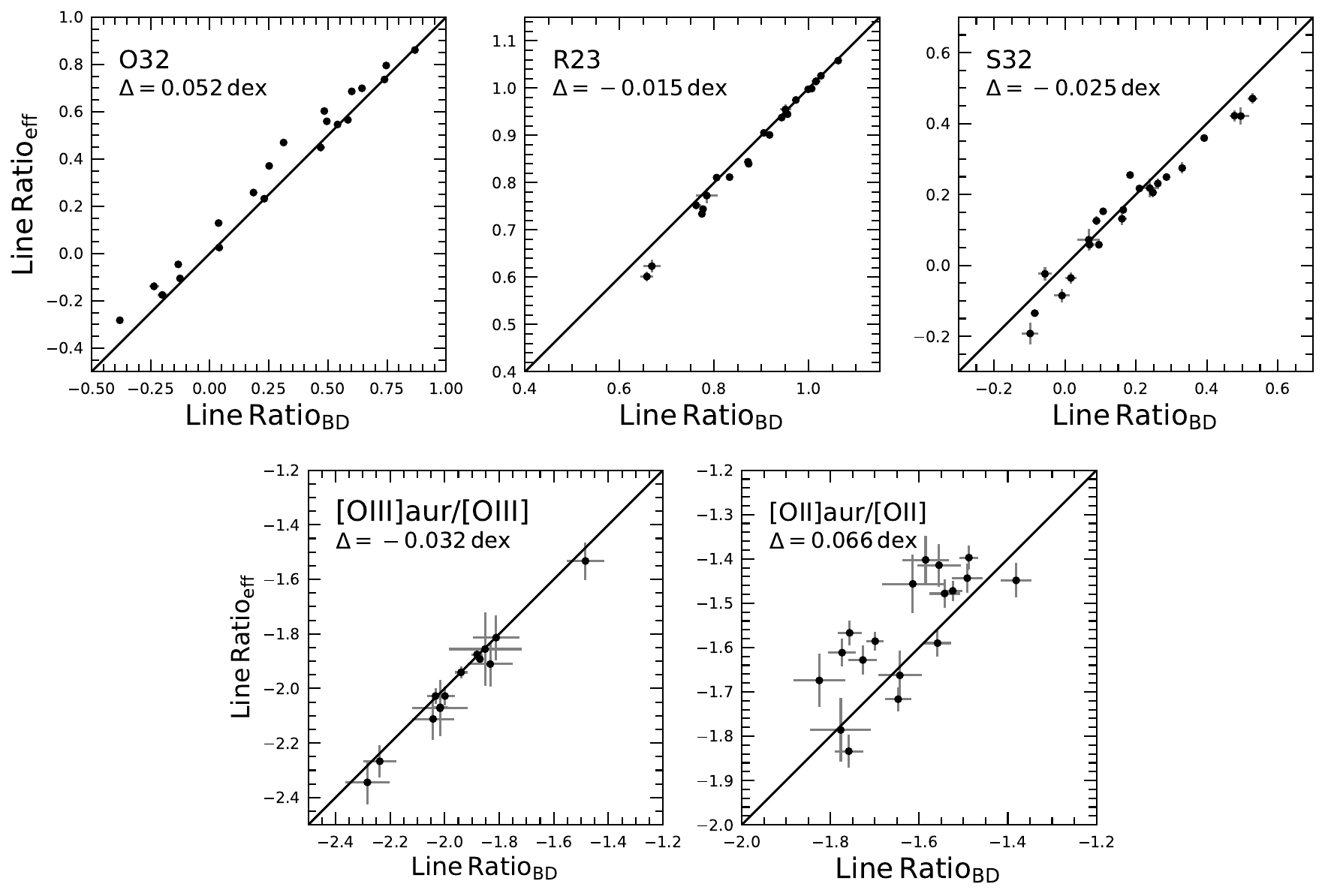}
    \caption{Comparison of dust-corrected line ratios of O32, R23, and
      S23, defined in Equations~\ref{eq:o32}, \ref{eq:r23}, and
      \ref{eq:s32}, assuming the individual effective nebular
      attenuation curves (${\rm Line\,Ratio}_{\rm eff}$) and the
      Balmer decrement and the Galactic extinction curve (${\rm
        Line\,Ratio}_{\rm BD}$).  The bottom two panels indicate the
      same for the auroral line ratios $\oiii\lambda 4364/\oiii\lambda
      5008$ and $\oii\lambda\lambda 7320, 7330/\oii\lambda\lambda
      3727,3730$.  The line of equality is indicated in black.  Each
      panel indicates the mean offset (e.g., ${\rm O32}_{\rm eff} -
      {\rm O32}_{\rm BD}$) in dex.}
   \label{fig:ratiocompare}
\end{figure*}

Figure~\ref{fig:ratiocompare} compares the dust-corrected line ratios
of O32, R23, S23, and the auroral line ratios $\oiii\lambda
4364/\oiii\lambda 5008$ and $\oii\lambda\lambda 7320,
7330/\oii\lambda\lambda 3727,3730$, assuming the two methods of
correcting for dust: either the individual nebular attenuation curves
(shown on the y-axis for each panel) or combining the Balmer decrement
with the Galactic extinction curve (shown on the x-axis for each
panel).  The two different methods of correcting for dust do not appreciably affect
the line ratios, all of which are typically shifted by $\la 0.07$\,dex on average, while the ordering of galaxies is generally preserved.  See Section~\ref{sec:effectonlineratios} for further discussion.

\section{Relationship between $\sfrneb$ and the Reddening, Dust Covering Fraction, and Obscured Fraction of Light}
\label{sec:app2}

\begin{figure*}
  \epsscale{1.0}
  \includegraphics[width=1.0\linewidth]{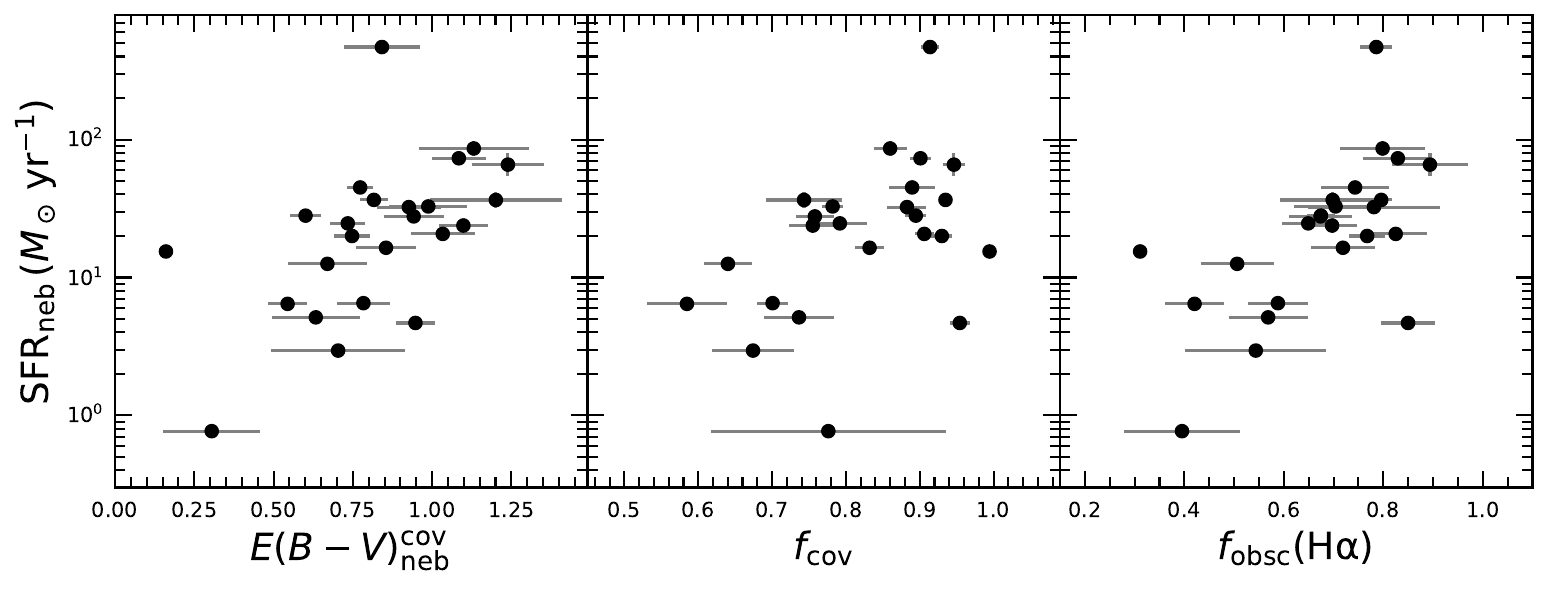}
    \caption{Correlations between $\sfrneb$ and $\ebmvneb^{\rm cov}$
      (left), $\fcov$ (middle), and $f_{\rm obsc}(\ha)/f_0(\ha)$ (right)
      derived from the covering-fraction model.}
   \label{fig:sfrtrends}
\end{figure*}

Figure~\ref{fig:sfrtrends}
presents the correlations between $\sfrneb$ and specific parameters
determined from the covering-fraction model: the line-of-sight
reddening, $\ebmvneb^{\rm cov}$; the covering fraction of dust,
$\fcov$; and the fraction of the intrinsic $\ha$ luminosity that
is obscured by dust, i.e., $f_{\rm obsc}(\ha)/f_0(\ha)$.

In the covering fraction model, 
\begin{eqnarray}
f_{\rm obsc}(\lambda) & = & \fcov f_0(\lambda)[1-10^{-0.4\ebmvneb^{\rm cov} k_{\rm neb}(\lambda)}]
\end{eqnarray} 
(see R26a for further details).  Not surprisingly, galaxies with
higher $\sfrneb$ tend to be redder, exhibit higher average $\fcov$,
and show a greater fraction of dust-obscured $\ha$ luminosity.  As
noted in Appendix~B of R26a, the uncertainties in $\ebmvneb^{\rm
  cov}$ and $\fcov$ are typically larger for galaxies with lower SFRs
as these galaxies have fewer detected ($S/N > 5$) higher-order Balmer
and Paschen lines.  Note that the quantities displayed in
Figure~\ref{fig:sfrtrends} are highly correlated with each other since
$\ebmvneb^{\rm cov}$, $\fcov$, and $f_{\rm obsc}(\ha)$ are all derived
from the same covering-fraction model which also constrains $\rv$ of
the effective attenuation curve (Section 5.3 of R26a) and hence
$\sfrneb$.  These ``correlations'' simply provide context for
understanding the relationship between total SFR and the distribution
and column density of dust.

\end{document}